\documentclass[a4paper, 10pt, final, runningheads]{svjour3}
\usepackage[a4paper, centering]{geometry}
\def\makeheadbox{}
\usepackage{color}
\definecolor{linkcol}{rgb}{0,0,0.4} 
\definecolor{citecol}{rgb}{0.5,0,0}
\usepackage[a4paper,dvips]{hyperref}
\hypersetup
{
  bookmarksopen=true,
  breaklinks=true,
  hyperindex=true,
  pdfmenubar=true, %menubar shown
  pdfhighlight=/O, %effect of clicking on a link
  colorlinks=true, %couleurs sur les liens hypertextes
  pdfpagemode=None, %aucun mode de page
  pdfpagelayout=SinglePage, %ouverture en simple page
  pdffitwindow=true, %pages ouvertes entierement dans toute la fenetre
  linkcolor=linkcol, %couleur des liens hypertextes internes
  citecolor=citecol, %couleur des liens pour les citations
  urlcolor=linkcol %couleur des liens pour les url
}
\hypersetup
{
  pdftitle={Sequential design of computer experiments%
    for the estimation of a probability of failure},
  pdfauthor={Julien Bect, David Ginsbourger, Ling Li, %
    Victor Picheny and Emmanuel Vazquez}, 
  pdfsubject={},
  pdfkeywords={Computer experiments; Sequential design; %
    Gaussian processes; Probability of failure; %
    Stepwise uncertainty reduction}
}
\journalname{Statistics and Computing}

\usepackage[latin1]{inputenc}
\usepackage[authoryear,round]{natbib}
\usepackage{amsmath,amsfonts,amssymb,bm,bbm}
\usepackage{graphicx,psfrag,epsfig}
\usepackage{multirow,enumerate}
\graphicspath{{./figs/}}

\usepackage{dsfont}                % \mathds
\renewcommand{\mathbb}{\mathds}    % \mathds --> \mathbb

\usepackage{color}

\smartqed  % flush right qed marks, e.g. at end of proof

\newcommand   \Xset  {\mathbb{X}}
\newcommand   \XX    {\Xset}
\newcommand   \RR    {\mathbb{R}}

\newcommand   \Rplus {\RR_{+}}

\renewcommand \P   {\mathsf{P}}
\newcommand   \Q   {\mathsf{Q}}
\newcommand   \PX  {\P_{\Xset}}

\newcommand   \FF    {\mathcal{F}}
\newcommand   \Acal  {\mathcal{A}}
\newcommand   \Ical  {\mathcal{I}}
\newcommand   \Bcal  {\mathcal{B}}

\newcommand   \Ncal  {\mathcal{N}}

\newcommand \EE   {\mathsf{E}}

\DeclareMathOperator \cov  {cov}

\DeclareMathOperator \Span {{\rm span}}

\DeclareMathOperator* \argmin {argmin}
\DeclareMathOperator* \argmax {argmax}

\newcommand{\one}{{\mathbb{1}}}
\newcommand{\tr}{^{\mathsf{T}}}

\renewcommand{\dot}{\,\bm{\cdot}\,}

\providecommand{\abs}[1]{\lvert#1\rvert}

\newcommand \ddiff {\mathrm{d}}

\newcommand \dy    {\ddiff y}
\newcommand \dz    {\ddiff z}
\newcommand \du    {\ddiff u}

\renewcommand{\hat}{\widehat}

\newcommand \x   {\underline{x}}

\newcommand \Xv  {\underline{X\mskip -1.5mu} \mskip 1mu}

\newcommand \dPX   {\ddiff\PX}
\newcommand \ahat  {\hat{\alpha}}
\newcommand \xihat {\hat{\xi}}
\newcommand \sn    {\sigma_n}
\newcommand \GP    {\mathrm{GP}}
\newcommand \wtR   {\widetilde{R}}
\newcommand \thres {u}
\newcommand \se    {\sigma_\varepsilon}

\newcommand \XvN[1][]  {\underline{X\mskip -2mu}_N^{\mskip 1mu#1}\mskip 1mu}

\newcommand \JSUR[1]   {J_{#1}^{\text{\tiny SUR}}}
\newcommand \JTIMSE[1] {J_{#1}^{\text{\tiny tIMSE}}}
\newcommand \JECH[1]   {J_{#1}^{\text{\tiny EGL}}}
\newcommand \JRB[1]    {J_{#1}^{\text{\tiny RB}}}

\newcommand \normPDF \varphi
\newcommand \normCDF \Phi

\newcommand{\bmid}{\;\big\lvert\;}

\begin{document}

\sloppy

\title{Sequential design of computer experiments for the estimation of
  a probability of failure}

\author{Julien Bect$^{\star}$  \and David
  Ginsbourger \and Ling~Li \and Victor Picheny \and Emmanuel Vazquez$^{\star}$ }

%%%%%%%%%%%% CUSTOM FOR POSTPRINT %%%%%%%%%%%%%%%%%%%%%%%%%%%%%
\authorrunning{%
  Author-generated postprint version. %
  See \href{http://dx.doi.org/10.1007/s11222-011-9241-4}%
  {DOI:10.1007/s11222-011-9241-4} for the published version.}
%%%%%%%%%%%%%%%%%%%%%%%%%%%%%%%%%%%%%%%%%%%%%%%%%%%%%%%%%%%%%%%

\institute{$^{\star}$ corresponding authors\\[1ex]
  J. Bect, L. Li, and E. Vazquez \at
  SUPELEC, Gif-sur-Yvette, France.\\
  \email{ \{firstname\}.\{lastname\}@supelec.fr}
  \and V. Picheny \at Ecole Centrale Paris, Chatenay-Malabry, France.\\
  \email{victor.picheny@ecp.fr}
  \and  D. Ginsbourger \at
Institute of Mathematical Statistics and Actuarial Science,
University of Bern, Switzerland.\\
\email{david.ginsbourger@stat.unibe.ch}
}

\date{Received: date / Accepted: date}

\maketitle

\begin{abstract}
  This paper deals with the problem of estimating the volume of the
  excursion set of a function $f:\RR^d\to\RR$
  above a given threshold, under a probability measure on $\RR^d$ that
  is assumed to be known. In the industrial world, this corresponds to
  the problem of estimating a probability of failure of a system. When
  only an expensive-to-simulate model of the system is available, the
  budget for simulations is usually severely limited and therefore
  classical Monte Carlo methods ought to be avoided. One of the main
  contributions of this article is to derive \textsl{SUR}
  (\emph{stepwise uncertainty reduction}) strategies from a
  Bayesian-theoretic formulation of the problem of estimating a
  probability of failure. These sequential strategies use a Gaussian
  process model of $f$ and aim at performing evaluations of $f$ as
  efficiently as possible to infer the value of the probability of
  failure. We compare these strategies to other strategies also based on
  a Gaussian process model for estimating a probability of failure.

\end{abstract}

\section{Introduction}

The design of a system or a technological product has to take into
account the fact that some design parameters are subject to unknown
variations that may affect the reliability of the system.  In
particular, it is important to estimate the probability of the system
to work under abnormal or dangerous operating conditions due to 
random dispersions of its characteristic parameters.  The
\emph{probability of failure} of a system is usually expressed as the
probability of the excursion set of a function above a fixed
threshold. More precisely, let $f$ be a measurable real function defined
over a probability space $(\XX,\Bcal({\XX}),\P_{\XX})$, with
$\XX\subseteq \RR^{d}$, and let $\thres \in \RR$ be a threshold.  The problem to
be considered in this paper is the estimation of the volume, under
$\P_{\XX}$, of the excursion set
\begin{equation}
  \label{eq:1}
  \Gamma := \{x \in \XX : f(x) > \thres\}
\end{equation}
of the function $f$ above the level $\thres$. In the context of robust
design, the volume $\alpha:=\P_{\XX}(\Gamma)$ can be viewed as the
probability of failure of a system: the probability $\P_{\XX}$ models
the uncertainty on the input vector $x \in \XX$ of the system---the
components of which are sometimes called \emph{design variables} or
\emph{factors}---and $f$ is some deterministic performance function
derived from the outputs of a deterministic model of the
system\footnote{Stochastic simulators are also of considerable
  practical interest, but raise specific modeling and computational
  issues that will not be considered in this paper.}. The evaluation
of the outputs of the model for a given set of input factors may
involve complex and time-consuming computer simulations, which turns
$f$ into an expensive-to-evaluate function. When $f$ is expensive to
evaluate, the estimation of $\alpha$ must be carried out with a
\emph{restricted number of evaluations} of $f$, generally excluding
the estimation of the probability of excursion by a Monte Carlo
approach. Indeed, consider the empirical estimator
\begin{equation}
  \label{eq:2}
  \alpha_m := \frac{1}{m}\sum_{i=1}^m \one_{\{f(X_i) > \thres \}}\,,
\end{equation}
where the $X_i$s are independent random variables with distribution
$\P_{\XX}$. According to the strong law of large numbers, the
estimator~$\alpha_m$ converges to $\alpha$ almost surely when $m$
increases. Moreover, it is an unbiased estimator of $\alpha$, i.e.
$\EE(\alpha_m)=\alpha$. Its mean square error is
$$
\EE\big((\alpha_m - \alpha)^2 \big) =
\frac{1}{m}\, \alpha\big( 1- \alpha\big)\,.
$$
If the probability of failure $\alpha$ is small, then the standard
deviation of $\alpha_m$ is approximately $\sqrt{\alpha/m}$. To
achieve a given standard deviation $\delta \alpha$ thus requires
approximately $1/({\delta^2 \alpha})$ evaluations, which can be
prohibitively high if $\alpha$ is small. By way of illustration, if
$\alpha=2\times10^{-3}$ and $\delta=0.1$, we obtain $m=50000$.  If one
evaluation of~$f$ takes, say, one minute, then the entire estimation
procedure will take about 35 days to complete. Of course, a host of 
refined random sampling methods have been proposed to improve over the basic
Monte Carlo convergence rate; for instance, methods based on importance sampling with
cross-entropy \citep{rubinstein:2004:ce},
subset sampling \citep{au01:_estim} or line sampling
\citep{Pradlwarter2007208}.  They will not be considered here for the
sake of brevity and because the required number of function evaluations
is still very high.

Until recently, all the methods that do not require a large number of
evaluations of~$f$ were based on the use of parametric approximations
for either the function~$f$ itself or the boundary $\partial \Gamma$
of $\Gamma$. The so-called response surface method falls in the first
category \citep[see, e.g.,][and references therein]{bucher:1990,
  raja:1993:newlook}. The most popular approaches in the second
category are the first- and second-order reliability method (FORM and
SORM), which are based on a linear or quadratic approximation
of~$\partial\Gamma$ around the \emph{most probable failure point}
\citep[see, e.g.,][]{bjerager90:_comput_method_for_struc_reliab_analy}.  
In all these methods, the accuracy of the estimator depends on the
actual shape of either~$f$ or~$\partial \Gamma$ and its resemblance to
the approximant: they do not provide statistically consistent
estimators of the probability of failure.

This paper focuses on sequential sampling strategies based on Gaussian
processes and kriging, which can been seen as a \emph{non-parametric}
approximation method. Several strategies of this kind have been
proposed recently in the literature by \citet{ranjan:2008},
\citet{bichon:2008}, \citet{piche:08:timse} and
\citet{echard:2010, echard:2010:b}. The idea is that the Gaussian process model,
which captures prior knowledge about the unknown function~$f$, makes
it possible to assess the uncertainty about the position of~$\Gamma$
given a set of evaluation results. This line of research has its roots
in the field of design and analysis of computer experiments
\citep[see, e.g.,][]{sac89, Currin91, Welch92, oakley:2002:bayesian,
  oakley:2004:probabilistic, oakley:2004:perc,
  bayarri:2007:fvcm}. More specifically, kriging-based sequential
strategies for the estimation of a probability of failure are closely
related to the field of Bayesian global optimization \citep{Mockus,
  MockusB, Jones, villemonteix:2008:phd, villemonteix:2009:iago,
  GinsbPhd}.

The contribution of this paper is twofold.  First, we introduce a
Bayesian decision-theoretic framework from which the theoretical form
of an optimal strategy for the estimation of a probability of failure
can be derived.  One-step lookahead sub-optimal
strategies are then proposed\footnote{Preliminary accounts of this work have been
  presented in \citet{vazquez:2007:ds} and~\citet{vaz:09:sysid}.}, which are
suitable for numerical evaluation and implementation on
computers. These strategies will be called SUR (stepwise uncertainty
reduction) strategies in reference to the work of D. Geman and its 
collaborators \citep[see, e.g.][]{fleuret:1999:sur}. Second,
we provide a review in a unified framework of all the kriging-based
strategies proposed so far in the literature and compare them
numerically with the SUR strategies proposed in this paper.

The outline of the paper is as follows. Section~\ref{sec:BayesDTF}
introduces the Bayesian framework and recalls the basics of dynamic
programming and Gaussian processes. Section~\ref{sec:SUR} introduces
SUR strategies, from the decision-theoretic underpinnings, down to the
implementation level. Section~\ref{sec:otherstrat} provides a review
of other kriging-based strategies proposed in the
literature. Section~\ref{sec:numerical} provides some illustrations
and reports an empirical comparison of these sampling
criteria. Finally, Section~\ref{sec:discuss} presents conclusions and
offers perspectives for future work.

\section{Bayesian decision-theoretic framework}
\label{sec:BayesDTF}

\subsection{Bayes risk and sequential strategies}
\label{sec:loss-function-bayes}

Let $f$ be a continuous function. We shall assume that $f$ corresponds
to a computer program whose output is not a closed-form expression of
the inputs.  Our objective is to obtain a numerical approximation of the
probability of failure
\begin{equation}
  \label{eq:3}
  \alpha(f) = \PX\{ x\in\XX: f(x) > \thres\} = \int_{\XX} \one_{f>\thres}\, \dPX\,,
\end{equation}
where $\one_{f>u}$ stands for the characteristic function of the
excursion set $\Gamma$, such that for any $x\in\XX$, $\one_{f>\thres}(x)$
equals one if $x\in\Gamma$ and zero otherwise. 
The approximation of $\alpha(f)$ has to be built from
a set of computer experiments, where an experiment simply consists in
choosing an $x\in\XX$ and computing the value of $f$ at $x$.  The
result of a pointwise evaluation of $f$ carries information about $f$
and quantities depending on $f$ and, in particular, about $\one_{f>\thres}$
and $\alpha(f)$. In the context of expensive computer experiments, we
shall also suppose that the number of evaluations is limited. Thus,
the estimation of $\alpha(f)$ must be carried out using a fixed
number, say $N$, of evaluations of~$f$.

A sequential non-randomized algorithm to estimate~$\alpha(f)$ with a budget of~$N$
evaluations is a pair~$\left( \XvN, \ahat_N \right)$,
\begin{equation*}
  \XvN : f  \mapsto \XvN(f) = (X_1(f), X_2(f),\ldots,
  X_N(f))\in\XX^N\,, \qquad
  \ahat_N : f \mapsto \ahat_N(f) \in \Rplus \,,
\end{equation*}
with the following properties: 
\begin{enumerate}[a)]
\item There exists $x_1 \in \XX$ such that $X_1(f) = x_1$, i.e. $X_1$
  does not depend on~$f$.
\item Let $Z_n(f) = f(X_n(f))$, $1 \le n \le N$. For all $1 \leq n <
  N$, $X_{n+1}(f)$ depends measurably\footnote{i.e., there is a
    measurable map $\varphi_n:\left( \XX \times \RR \right)^n \to \XX$
    such that $X_n = \varphi_n \circ \Ical_n$} on $\Ical_{n}(f)$,
  where $\Ical_{n} = \left( \left(X_1, Z_1 \right), \ldots, \left(
      X_n, Z_n \right) \right)$.
\item $\ahat_N(f)$ depends measurably on~$\Ical_N(f)$.
\end{enumerate}
The mapping $\XvN$ will be  referred to as a strategy, or policy, or
design of experiments, and $\ahat_{N}$ will be called an estimator.  The
algorithm~$\left( \XvN, \ahat_N \right)$ describes a sequence of
decisions, made from an increasing amount of information:
$X_1(f) = x_1$ is chosen prior to any evaluation; for each
$n=1,\ldots, N-1$, the algorithm uses  information
$\Ical_{n}(f)$ to choose the next evaluation
point~$X_{n+1}(f)$; the estimation $\ahat_N(f)$ of~$\alpha(f)$ is
the terminal decision. In some applications, the class of sequential
algorithms must be further restricted: for instance, when $K$ computer
simulations can be run in parallel, algorithms  that query batches
of~$K$ evaluations at a time may be preferred \citep[see, e.g.][]{ginsbourger:2010:kws}.
In this paper no such restriction is imposed.

The choice of the estimator~$\ahat_N$ will be addressed in
Section~\ref{sec:estimators}: for now, we simply assume that an
estimator has been chosen, and focus on the problem of finding a good
strategy~$\XvN$; that is, one that will produce a good final
approximation $\ahat_N(f)$ of $\alpha(f)$. Let
$\Acal_N$ be the class of all strategies~$\XvN$
that query sequentially $N$ evaluations of $f$. Given a loss function $L:\RR \times
\RR \to \RR$, we define the error of approximation of a strategy
$\XvN\in\Acal_N$ on~$f$ as $\epsilon(\XvN,f) = L(\ahat_N(f),
\alpha(f))$. In this paper, we shall consider the quadratic loss
function, so that $ \epsilon(\XvN,f) = (\ahat_N(f) -\alpha(f))^2$.

We adopt a Bayesian approach to this decision problem: the unknown
function~$f$ is considered as a sample path of a real-valued random
process $\xi$ defined on some probability space $(\Omega, \Bcal,
\P_0)$ with parameter in $x\in\XX$, and a good strategy is a strategy
that achieves, or gets close to, the \emph{Bayes risk} $r_{\rm B} :=
\inf_{\XvN \in\Acal_N} \EE_0\left( \epsilon(\XvN,\xi) \right)$,
where~$\EE_0$ denotes the expectation with respect to~$\P_0$. From a
subjective Bayesian point of view, the stochastic model~$\xi$ is a
representation of our uncertain initial knowledge about~$f$. From a more
pragmatic perspective, the prior distribution can be seen as a tool to
define a notion of a good strategy in an average sense. Another
interesting route, not followed in this paper, would have been to
consider the minimax risk $\inf_{\XvN \in\Acal_N} \max_f \EE_0\left(
  \epsilon(\XvN,\xi) \right)$ over some class of functions.

\textbf{Notations.} From now on, we shall consider the stochastic
model~$\xi$ instead of the deterministic function~$f$ and, for
abbreviation, the explicit dependence on~$\xi$ will be dropped when no
there is no risk of confusion; e.g., $\ahat_N$ will denote the random
variable $\ahat_N (\xi)$, $X_n$ will denote the random variable
$X_n(\xi)$, etc.  We will use the notations~$\FF_n$, $\P_n$ and~$\EE_n$
to denote respectively the $\sigma$-algebra generated by~$\Ical_n$, the
conditional distribution $\P_0\left( \dot \mid \FF_n \right)$ and the
conditional expectation $\EE_0\left( \dot \mid \FF_n\right)$. Note that
the dependence of~$X_{n+1}$ on~$\Ical_n$ can be rephrased by saying that
$X_{n+1}$ is $\FF_n$-measurable. Recall that $\EE_n\left( Z \right)$ is
$\FF_n$-measurable, and thus can be seen as a measurable function
of~$\Ical_n$, for any random variable~$Z$.

\subsection{Optimal and $k$-step lookahead strategies}
\label{sec:optim_strategies}

It is well-known \citep[see, e.g.,][]{berry:1985:bp, MockusB, bertsekas:1995:dpoc1} 
that an optimal strategy for such a finite horizon problem\footnote{in
  other words, a sequential decision problem where the total number of steps
  to be performed is known from the start}, i.e. a strategy~$\XvN[\star]
\in \Acal_N$ such that $\EE_0\left( \epsilon(\XvN[\star],\xi) \right) = r_{\rm
  B}$, can be formally obtained by \emph{dynamic programming}: let $R_N =
\EE_N\left( \epsilon(\XvN,\xi) \right) = \EE_N\big( (\ahat_N -
\alpha)^2 \big)$ denote the terminal risk and define by backward induction
\begin{equation}
  \label{equ:dp:1}
  R_n = \displaystyle \min_{x \in \XX}~
  \EE_n\big( R_{n+1} \mid X_{n+1}=x \big)\,,
  \quad n=N-1,\ldots, 0.
\end{equation}
To get an insight into~(\ref{equ:dp:1}), notice that $R_{n+1}$,
$n=0,\ldots,N-1$, depends measurably on
$\Ical_{n+1} = (\Ical_{n}, X_{n+1}, Z_{n+1})$, so that
$\EE_n\big( R_{n+1} \mid X_{n+1}=x \big)$ is in fact an expectation with respect to
$Z_{n+1}$, and $R_n$ is an $\FF_n$-measurable random
variable.  Then, we have $R_0 = r_{\rm B}$, and the
strategy $\XvN[\star]$ defined by
\begin{equation}
  \label{equ:dp:2}
  X_{n+1}^{\star} = \argmin_{x \in \XX} \EE_n
  \big( R_{n+1}  \mid X_{n+1}=x \big)\,,\quad n=1,\ldots, N-1, 
\end{equation}
is optimal\footnote{Proving rigorously that, for a given~$\P_0$
  and~$\ahat_N$, equations~\eqref{equ:dp:1}
  and~\eqref{equ:dp:2} actually define a (measurable!) strategy
  $\XvN[\star] \in \Acal_N$ is technical problem that is not of
  primary interest in this paper. This can be done for instance, in
  the case of a Gaussian process with continuous covariance function
  (as considered later), by proving that $x \mapsto \EE_n\left(
  R_{n+1} \mid X_{n+1}(\xi)=x \right)$ is a continuous function
  on~$\XX$ and then using a measurable selection theorem.}. It is
crucial to observe here that, for this dynamic programming problem,
both the space of possible actions and the space of possible outcomes
at each step are continuous, and the state space $\left( \XX \times
  \RR \right)^n$ at step~$n$ is of dimension $n(d+1)$. Any direct
attempt at solving \eqref{equ:dp:1}--\eqref{equ:dp:2}
numerically, over an horizon $N$ of more than a few steps, will suffer
from the curse of dimensionality.

Using \eqref{equ:dp:1}, the optimal strategy can be expanded as
\begin{equation*}
  X^{\star}_{n+1} = \argmin_{x \in \XX}\, \EE_n\left(
  \min_{X_{n+2}}\, \EE_{n+1}\, \ldots\,
  \min_{X_N}\, \EE_{N-1}\, R_N 
  \Bigm| X_{n+1} = x \right) \,.
\end{equation*}
A very general approach to construct sub-optimal---but hopefully
good---strategies is to truncate this expansion after $k$ terms,
replacing the exact risk~$R_{n+k}$ by any available
surrogate~$\wtR_{n+k}$. Examples of such surrogates will be given in
Sections~\ref{sec:SUR} and~\ref{sec:otherstrat}. The resulting
strategy,
\begin{equation}
  \label{equ:strat-ksl}
  X_{n+1} = \argmin_{x \in \XX}\, \EE_n\left(
  \min_{X_{n+2}}\, \EE_{n+1}\, \ldots\,
  \min_{X_{n+k}}\, \EE_{n+k-1}\, \wtR_{n+k}
  \Bigm| X_{n+1} = x \right) \,.
\end{equation}
is called a \emph{$k$-step lookahead strategy} \citep[see,
e.g.,][Section~6.3]{bertsekas:1995:dpoc1}. Note that both the optimal
strategy~\eqref{equ:dp:2} and the $k$-step lookahead strategy
implicitly define a \emph{sampling criterion} $J_n(x)$,
$\FF_{n}$-measurable, the minimum of which indicates the next
evaluation to be performed. For instance, in the case of the $k$-step
lookahead strategy, the sampling criterion is
\begin{equation*}
  J_n(x) = \EE_n\left(
    \min_{X_{n+2}}\, \EE_{n+1}\, \ldots\,
    \min_{X_{n+k}}\, \EE_{n+k-1}\, \wtR_{n+k}
    \Bigm| X_{n+1} = x \right) \,.
\end{equation*}
In the rest of the paper, we restrict our attention to the class of
one-step lookahead strategies, which is, as we shall see in
Section~\ref{sec:SUR}, large enough to provide very efficient
algorithms. We leave aside the interesting question of whether more
complex $k$-step lookahead strategies (with $k \ge 2$) could provide
a significant improvement over the strategies examined in this paper.

\begin{remark}
  In practice, the analysis of a computer code usually begins with an
  exploratory phase, during which the output of the code is computed on a
  \emph{space-filling design} of size $n_0 < N$ \cite[see,
  e.g.,][]{santner:2003:dace}. Such an exploratory phase will be
  colloquially 
  referred to as the \emph{initial design}. Sequential strategies such
  as~\eqref{equ:dp:2} and~\eqref{equ:strat-ksl} are meant to be used
  after this initial design, at steps $n_0+1$, \ldots, $N$. An
  important (and largely open) question is the choice of the
  size~$n_0$ of the initial design, for a given global budget~$N$. As
  a rule of thumb, some authors recommend to start with a sample size
  proportional to the dimension~$d$ of the input space~$\XX$, for
  instance $n_0 = 10\, d$ ; see \citet{loeppky:2009:css} and the
  references therein.
\end{remark}

\subsection{Gaussian process priors}
\label{sec:gauss-proc-priors}

Restricting $\xi$ to be a Gaussian process makes it possible to deal
with the conditional distributions~$\P_n$ and conditional expectations
$\EE_n$ that appear in the strategies above. The idea of modeling an
unknown function~$f$ by a Gaussian process has originally
been introduced circa 1960 in 
time series analysis \citep{parzen62}, optimization theory
\citep{kushner64} and geostatistics \citep[see, e.g.,][and the
references therein]{chiles99}. Today, 
the Gaussian process model plays a central role in the design and analysis of
computer experiments \citep[see,
e.g.,][]{sac89,Currin91,Welch92,santner:2003:dace}. Recall that the
distribution of a Gaussian process~$\xi$ is uniquely determined by its
mean function $m(x) := \EE_0(\xi(x))$, $x\in\XX$, and its covariance
function $k(x,y) := \EE_0\left( (\xi(x) - m(x))(\xi(y) - m(y)) \right)$,
$x,y\in\XX$. Hereafter, we shall use the notation $\xi \sim \GP\left(
  m,\, k \right)$ to say that $\xi$ is a Gaussian process with
mean function $m$ and covariance function  $k$.

Let $\xi \,\sim\, \GP\left( 0,\, k \right)$ be a zero-mean Gaussian
process. The best linear unbiased predictor (BLUP) of $\xi(x)$ from observations $\xi(x_i)$,
$i=1,\ldots, n$, also called the \emph{kriging predictor} of $\xi(x)$,
is the orthogonal projection
\begin{equation}
  \label{eq:def-krig-pred}
  \xihat( x; \x_n ) \;:=\; \sum_{i=1}^n \lambda_i(x;\x_n)\, \xi(x_i)  
\end{equation}
of $\xi(x)$ onto $\Span\{\xi(x_i),i=1,\ldots ,n\}$.  Here, the notation $\x_n$
stands for the set of points $\x_n = \{x_1,\ldots, x_n\}$. The weights
$\lambda_i(x;\x_n)$ are the solutions of a system of linear equations
\begin{equation}
\label{eq:krigsys}
      k(\x_n,\x_n) {\lambda}(x;\x_n) =  {{k}} (x,\x_n)
\end{equation}
where $k(\x_n,\x_n)$ stands for the $n\times n$ covariance matrix of the
observation vector, $\lambda(x;\x_n)=(\lambda_1(x;\x_n),\ldots,
\lambda_n(x;\x_n))\tr$, and $k(x,\x_n)$ is a vector with entries
$k(x,x_i)$. The function $x\mapsto \xihat( x; \x_n )$ conditioned on
$\xi(x_1)=f(x_1),\ldots, \xi(x_n) = f(x_n)$, is deterministic, and
provides a cheap \emph{surrogate model} for the true function~$f$
\citep[see, e.g.,][]{santner:2003:dace}. The covariance function of the
error of prediction, also called \emph{kriging covariance} is given by
\begin{align}
  \label{eq:def-krig-cov}
  k(x,y; \x_n)
  \;&:=\; \cov \left( \xi(x)- \xihat(x;\x_n), \xi(y) - \xihat(y;\x_n)
  \right) \nonumber \\
  &=\; k(x,y) - \sum_i \lambda_i(x;\x_n) \, k(y,x_i) \,.
\end{align}
The variance of the prediction error, also called the \emph{kriging
  variance}, is defined as $ \sigma^2(x; \x_n) = k(x,x; \x_n)$.
One fundamental property of a zero-mean Gaussian process is the
following \citep[see, e.g.,][Chapter~3]{chiles99} :
\begin{proposition}
  If $\xi \,\sim\, \GP\left(0,\, k \right)$, then the random process
  $\xi$ conditioned on the $\sigma$-algebra $\FF_{n}$ generated by
  $\xi(x_1),\ldots,\xi(x_n)$, which we shall denote by $\xi\mid\FF_n$,
  is a Gaussian process with mean~$\xihat(\dot;\,\x_n)$ given
  by~\eqref{eq:def-krig-pred}-\eqref{eq:krigsys} and covariance $k\left(\, \dot,
    \dot ; \, \x_{n} \right)$ given by~\eqref{eq:def-krig-cov}.  In
  particular, $\xihat(x;\x_n) = \EE_0\bigl( \xi(x) \mid \FF_{n}
  \bigr)$ is the best $\FF_n$-measurable predictor of~$\xi(x)$, for all
  $x\in\XX$.
\end{proposition}

In the domain of computer experiments, the mean of a Gaussian process is
generally written as a linear parametric function
\begin{equation}
\label{eq:4}
  m(\dot) = \beta\tr {h}(\dot)\,,  
\end{equation}
where $\beta$ is a vector of unknown parameters, and $h = \left( h_1,
  \ldots, h_l \right)\tr$ is an $l$-dimensional vector of functions
(in practice, polynomials). The simplest case is when the mean
function is assumed to be an unknown constant $m$, in which case we
can take $\beta=m$ and $h: x \in \XX \mapsto 1$. The covariance
function is generally written as a translation-invariant function:
$$
k: (x,y)\in\XX^2 \mapsto \sigma^2\, \rho_\theta(x-y)\,,
$$
where $\sigma^2$ is the variance of the (stationary) Gaussian process
and $\rho_{\theta}$ is the correlation function, which generally depends on
a  parameter vector $\theta$. When the mean is written under the
form~(\ref{eq:4}), the kriging predictor is again a linear combination
of the observations, as in
(\ref{eq:def-krig-pred}), and the weights $\lambda_i(x;\x_n)$ are again
solutions  of a system of linear equations \citep[see,
e.g.,][]{chiles99}, which can be written under a matrix form as
\begin{equation}
\label{eq:mat_sol}
  \left(  \begin{array}{cc}
      k(\x_n,\x_n)  & {h(\x_n)}\tr \\
      {h(\x_n)}  & 0
    \end{array}
  \right) 
  \left(  \begin{array}{c}
      {\lambda}(x;\x_n) \\
      {\mu}(x)
    \end{array}
  \right)
  =
  \left(  \begin{array}{c}
      {{k}} (x,\x_n)\\
      {{h}} (x)
    \end{array}
  \right)\,,
\end{equation}
where $h(\x_n)$ is an $l\times n$ matrix with entries $h_i(x_j)$, $i =
1, \ldots, l$, $j = 1, \ldots, n$, $\mu$ is a vector of Lagrange
coefficients ($k(\x_n,\x_n)$, $\lambda(x;\x_n)$, $k(x,\x_n)$ as
above). The kriging covariance function is given in this case by
\begin{align}
  \label{eq:def-krig-cov2}
  k(x,y; \x_n)
  \;&:=\; \cov \left( \xi(x)- \xihat(x;\x_n), \xi(y) - \xihat(y;\x_n)
  \right) \nonumber \\
  &=\; k(x,y) - \lambda(x;\x_n)\tr \, k(y,\x_n) - \mu(x)\tr h(y) \,.
\end{align}
The following result holds \citep{kilmeldorf70, ohagan:1978:curvefitting}:
\begin{proposition}
\label{prop:gauss-proc-priors}
Let $k$ be a covariance function.
\begin{equation*}
  \text{If }\; 
  \left\{
      \begin{aligned}
        &\xi \mid m  \;\sim\; \GP\left( m,\, k \right)\\
        &m:x\mapsto \beta\tr h(x),\, \beta  \;\sim\; \mathcal{U}_{\RR^l}
      \end{aligned}      
  \right.
  \;\text{ then }\;
    \xi \mid \FF_n \,\sim\, \GP\left( \xihat(\dot;\x_n),\, k(\dot,\dot;\,\x_{n}) \right)\,,
\end{equation*}
where $\mathcal{U}_{\RR^l}$ stands for the (improper) uniform
distribution over $\RR^l$, and where $\xihat(\dot;\x_n)$
and~$k(\dot,\dot;\,\x_{n})$ are given
by~\eqref{eq:def-krig-pred},~\eqref{eq:mat_sol} and~\eqref{eq:def-krig-cov2}.
\end{proposition}
Proposition~\ref{prop:gauss-proc-priors} justifies the use of kriging in
a Bayesian framework provided that the covariance function of $\xi$ is
known. However, the covariance function is rarely assumed to be known in
applications. Instead, the covariance
function is generally taken in some parametric class (in this paper, we use the
so-called Mat\'ern covariance function, see Appendix~\ref{sec:matern}).
A \emph{fully Bayesian} approach also
requires to choose a prior distribution for the unknown parameters of
the covariance \citep[see,
e.g.,][]{handcock:1993, kennedy:2001:bccm, paulo:2005:dpgp}.
Sampling techniques (Monte Carlo Markov Chains,
Sequential Monte Carlo...) are then generally used to approximate the posterior
distribution of the unknown covariance parameters. 
Very often, the popular \emph{empirical Bayes} approach is used
instead, which consists in plugging-in the maximum likelihood (ML)
estimate to approximate the posterior distribution of~$\xi$. This
approach has been used in previous papers about contour estimation  or probability
of failure estimation \citep{piche:08:timse, ranjan:2008,
  bichon:2008}. In Section~\ref{sec:num:structural} we will adopt a
plug-in approach as well.

\textbf{Simplified notations.} In the rest of the paper, we shall use
the following simplified notations when there is no risk of confusion:
$\xihat_n(x) := \xihat(x;\Xv_n)$, $\sigma^2_n(x) :=
\sigma^2(x;\Xv_n)$.

\subsection{Estimators of the probability of failure}
\label{sec:estimators}

Given a random process $\xi$ and a strategy~$\Xv_N$, the optimal
estimator that  minimizes
$\EE_0\left( (\alpha - \ahat_n)^2\right)$ among all $\FF_n$-measurable
estimators $\ahat_n$, $1\leq n \leq N$, is 
\begin{equation}
  \label{eq:estimator1}
  \ahat_n \;=\; \EE_n\left( \alpha \right)
  \;=\; \EE_n \left( \int_{\XX} \one_{\xi  > u}\, \dPX \right) 
  \;=\; \int_\XX p_n\, \dPX \,,
\end{equation}
where 
\begin{equation}
  p_n:x\in\XX \mapsto \P_n \left\{ \xi(x) > u \right\}\,.
\end{equation}
When $\xi$ is a Gaussian process, the  probability $p_n(x)$ of
exceeding~$u$ at~$x\in \XX$ given $\Ical_n$ has a simple 
closed-form expression:
\begin{equation}
  p_n(x) \;=\; 
  1 \,-\, \normCDF\left( \frac{u - \xihat_n(x)}{\sn(x)} \right)
  \;=\; \normCDF\left( \frac{\xihat_n(x) - u}{\sn(x)} \right)\,,
\end{equation}
where $\normCDF$ is the cumulative distribution function of the normal
distribution. Thus, in the Gaussian case,  the estimator~\eqref{eq:estimator1} is amenable to a
numerical approximation, by integrating the excess
probability $p_n$ over $\XX$ (for instance using Monte Carlo sampling, 
see~Section~\ref{sec:SUR-discrete}).

Another natural way to obtain an estimator of $\alpha$ given $\Ical_n$
is to approximate the excess indicator $\one_{\xi > u}$ by a
hard classifier $\eta_n:\XX\to \{0,1\}$, where ``hard'' refers
to the fact that $\eta_n$ takes its values in~$\{ 0, 1 \}$. If
$\eta_n$ is close in some sense to $\one_{\xi>u}$, the estimator
\begin{equation}
  \label{equ:estim-pf:2a}
  \ahat_n = \int_{\XX} \eta_n \dPX
\end{equation}
should be close to $\alpha$.  More precisely, 
\begin{equation}
\label{eq:misclas}
\EE_n\left( (\ahat_n - \alpha)^2 \right) = \EE_n \left[ \left( \int (\eta_n -
  \one_{\xi>u})\dPX \right)^2 \right] \leq \int \EE_n\left((\eta_n -
  \one_{\xi>u})^2\right) \dPX\,.
\end{equation}
Let $\tau_n(x)= \P_n\{\eta_n(x) \neq \one_{\xi(x)>u} \} =
\EE_n\left((\eta_n(x) - \one_{\xi(x)>u})^2\right)$ be the probability of
misclassification; that is, the probability to predict a point above
(resp. under) the threshold when the true value is under (resp. above)
the threshold. Thus,~\eqref{eq:misclas} shows that it is desirable to
use a classifier $\eta_n$ such that $\tau_n$ is small for all
$x\in\XX$. For instance, the method called \textsc{smart}
\citep{deheeger:2007:svm} uses a support vector machine to build
$\eta_n$. Note that
\begin{equation*}
  \tau_{n(x)} \;=\;
  p_n(x)  \,+\, (1 - 2 p_n(x))\, \eta_n(x)\,. 
\end{equation*}
Therefore, the right-hand side of~\eqref{eq:misclas} is minimized 
if we set 
\begin{equation}
  \label{equ:estim-pf:2b}
  \eta_n(x) = \one_{p_n(x)>1/2} = \one_{\bar \xi_n(x)>u}\,,
\end{equation}
where $\bar\xi_n(x)$ denotes the posterior
median of $\xi(x)$. Then, we have $$\tau_n(x) = \min
(p_n(x),1-p_n(x)).$$ In the case of a Gaussian process, the posterior
median and the posterior mean are equal. Then, the classifier that
minimizes $\tau_n(x)$ for each $x\in\XX$ is  $\eta_n =
\one_{\xihat_n>u}$, in which case
\begin{equation}
  \tau_n(x) 
  \;=\; \P_n \left(
    (\xi(x) - \thres)(\xihat_n(x) - \thres) < 0 
  \right)
  \;=\; 1 \,-\, \normCDF\left( 
    \frac{\bigl| \xihat_n(x) - \thres \bigr|}{\sn(x)} 
  \right)\,.
\end{equation}
Notice that for $\eta_n = \one_{\xihat_n>u}$, we have $\ahat_n =
\alpha(\xihat_n)$. Therefore, this approach to obtain an estimator of
$\alpha$ can be seen as a type of plug-in estimation.

\bigbreak

\noindent \textbf{Standing assumption.}\ It will assumed in the rest
of the paper that~$\xi$ is a Gaussian process, or more generally that
$\xi \mid \FF_n \,\sim\, \GP\bigl( \xihat_n,\, k(\dot,\dot;\,\x_{n})
\bigr)$ for all $n \ge 1$ as in
Proposition~\ref{prop:gauss-proc-priors}.

\section{Stepwise uncertainty reduction}
\label{sec:SUR}

\subsection{Principle}
\label{sec:SUR:principle}

A very natural and straightforward way of building a one-step
lookahead strategy is to select \emph{greedily} each evaluation as if
it were the last one. This kind of strategy, sometimes called
\emph{myopic}, has been successfully applied in the field of Bayesian
global optimization \citep{Mockus, MockusB}, yielding the famous
\emph{expected improvement} criterion later popularized in the
Efficient Global Optimization (EGO) algorithm of \citet{Jones}. 

When the Bayesian risk provides a measure of the estimation error or
uncertainty (as in the present case), we call such a strategy a
\emph{stepwise uncertainty reduction} (SUR) strategy. In the field of
global optimization, the Informational Approach to Global Optimization
(IAGO) of \citet{villemonteix:2009:iago} is an example of a SUR
strategy, where the Shannon entropy of the minimizer is used instead
of the quadratic cost. When considered in terms of utility rather than
cost, such strategies have also been called \emph{knowledge gradient
  policies} by \citet{frazier:2008:kgp}.

Given a sequence of estimators $\left( \ahat_n \right)_{n \ge 1}$, a
direct application of the above principle using the quadratic loss
function yields the sampling criterion
\begin{equation}
  \label{eq:sur-1}
  J_n(x) = \EE_n \left( \left( \alpha - \ahat_{n+1} \right)^2 
    \mid X_{n+1} = x \right)\,.
\end{equation}
Having found no closed-form expression for this criterion, and no
efficient numerical procedure for its approximation, we will proceed
by upper-bounding and discretizing~\eqref{eq:sur-1} in order to get an
expression that will lend itself to a numerically tractable
approximation. By doing so, several SUR strategies will be
derived, depending on the choice of estimator (the posterior
mean~\eqref{eq:estimator1} or the plug-in
estimator~\eqref{equ:estim-pf:2a} with~\eqref{equ:estim-pf:2b}) and
bounding technique.

\subsection{Upper bounds of the SUR sampling criterion}
\label{sec:SUR:upper-bounds}

Recall that $\tau_n(x) = \min(p_n(x), 1-p_n(x))$ is the probability of
misclassification at~$x$ using the optimal
classifier~$\one_{\xihat_n(x) > u}$. Let us further denote by
$\nu_n(x) := p_n(x)\, \left( 1 - p_n(x) \right)$ the variance of the
excess indicator $\one_{\xi(x) \ge \thres}$.

\begin{figure}[htbp]
  \centering
  \psfrag{xxx}{$p_n$}
  \psfrag{yyy}{$\gamma_n$}
  \psfrag{aaaaaa}{\footnotesize $\gamma_n = p_n(1-p_n)$}
  \psfrag{bbbbbb}{\footnotesize $\gamma_n = \min(p_n,1-p_n)$}
  \includegraphics[width=7cm]{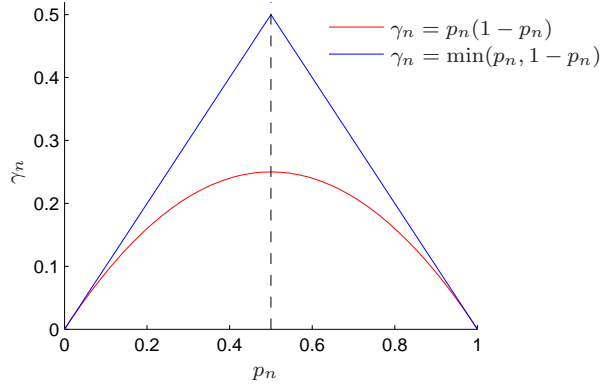}
  \caption{$\gamma_n$ as a function of~$p_n$ (see
    Proposition~\ref{prop:ub}). In both cases, $\gamma_n$ is maximum at~$p_n = 1/2$.}
  \label{fig:gamma-p}
\end{figure}

\begin{proposition}
  \label{prop:ub}
  Assume that either $\ahat_n = \EE_n\left( \alpha \right)$ or
  $\ahat_n = \int \one_{\xihat_n \ge u} \dPX$.  Define $G_n :=
  \int_\XX \sqrt{\gamma_n(y)}\dPX$ for all $n \in \{ 0, \ldots, N-1 \}$,
  with
  \begin{equation*}
    \gamma_n \;:=\;
    \begin{cases}
      \nu_n = p_n (1 - p_n) = \tau_n (1 - \tau_n) \,,
      & \text{if } \ahat_n = \EE_n\left( \alpha \right), \\
      \tau_n = \min(p_n, 1-p_n)\,,
      & \text{if } \ahat_n = \int \one_{\xihat_n \ge \thres} \dPX\,.
    \end{cases}
  \end{equation*}
  Then, for all $x \in \XX$ and all $n \in \{ 0, \ldots, N-1 \}$,
  \begin{equation*}
    J_n(x) \;\le\; \widetilde{J}_n(x)
    \;:=\; \EE_n\left( G_{n+1}^2 \mid X_{n+1} = x \right) \,.
  \end{equation*}
\end{proposition}
Note that $\gamma_n(x)$ is a function of~$p_n(x)$ that vanishes at~$0$
and~$1$, and reaches its maximum at~$1/2$; that is, when the
uncertainty on~$\one_{\xihat_n(x) > \thres}$ is maximal (see
Figure~\ref{fig:gamma-p}).

\proof First, observe that, for all $n \ge 0$, $\alpha - \ahat_n =
\int U_n\, \dPX$, with
\begin{equation}
  \label{proof:ub:equ1}
  U_n:x\in\XX \mapsto U_n(x) \;=\;
  \begin{cases}
    \one_{\xi(x) > \thres } -p_n(x)
    & \text{if } \ahat_n = \EE_n\left( \alpha \right), \\
    \one_{\xi(x) > \thres } -\one_{\xihat_n(x) > \thres}
       & \text{if } \ahat_n = \int \one_{\xihat_n \ge \thres} \dPX\,.
  \end{cases}
\end{equation}
Moreover, note that $\gamma_n = \left\lVert U_n \right\rVert^2_n$ in
both cases, where 
%$\lVert \dot\, \rVert_n$ denotes the (random)
%Hilbert norm in $L^2\left( \Omega, \Bcal, \P_n \right)$, i.e.,
 $\lVert \dot\, \rVert_n : L^2\left( \Omega, \Bcal, \P \right) \to L^2\left(
  \Omega, \FF_n, \P \right)$, $W \mapsto \EE_n\bigl( W^2 \bigr)^{1/2}$. 
Then, using the generalized Minkowski inequality \citep[see,
e.g.,][section~10.7]{vestrup:2003:meas-int} we get that 
\begin{equation}
  \label{equ:using-Minko}
  \Bigl\lVert \smallint U_n\, \dPX \Bigr\rVert_n 
  \;\le\;
  \int \lVert U_n \rVert_n\, \dPX
  \;=\; 
  \int \sqrt{\gamma_n}\, \dPX
  \;=\;
  G_n .  
\end{equation}
Finally, it follows from the tower property of conditional
expectations and~\eqref{equ:using-Minko} that, for all $n \ge 0$,
\begin{align*}
  J_n(x) 
  \;=\;& \EE_n \left(
    \lVert \alpha - \ahat_{n+1} \rVert_{n+1}^2 \mid X_{n+1} = x
  \right)
  \\
  \;=\;&
  \EE_n \left( 
    \bigl\lVert \smallint U_{n+1}\, \dPX \bigr\rVert_{n+1}^2 \Bigm| X_{n+1} = x 
  \right)
  \\
  \;\le\;&
  \EE_n \left(
    G_{n+1}^2 \mid X_{n+1} = x 
  \right)\,. 
\end{align*}
\qed

Note that two other upper-bounding sampling criteria readily follow from those
of Proposition~\ref{prop:ub}, by using the Cauchy-Schwarz inequality in
$L^2 \left( \XX, \Bcal(\XX), \PX \right)$:
\begin{equation}
  \label{equ:sur-ub-2}
  \widetilde{J}_n(x) \;\le\; \EE_n\left(
    \int \gamma_{n+1}\, \dPX  \Bigm| X_{n+1} = x \right) \,.
\end{equation}
As a result, we can write four SUR criteria, whose expressions are
summarized in Table~\ref{tab:expr-sur-crit}. Criterion $\JSUR{1,n}$
has been proposed in the PhD thesis of~\citet{pieramarti:2008} and in
conference papers \citep{vazquez:2007:ds, vaz:09:sysid}; the other
ones, to the best of our knowledge, are new. Each  criterion
is expressed as the conditional expectation of some (possibly squared)
$\FF_{n+1}$-measurable integral criterion, with an integrand that can
be expressed as a function of the probability of
misclassification~$\tau_{n+1}$. It is interesting to note that the
integral in~$\JSUR{4}$ is the integrated mean square error
(IMSE)\footnote{The IMSE criterion is usually applied to the response
  surface~$\xi$ itself \citep[see, e.g.,][]{box-draper:1987,
    sac89}. The originality here is to consider the IMSE of the
  process~$\one_{\xi > \thres}$ instead. Another way of adapting the
  IMSE criterion for the estimation of a probability of failure,
  proposed by \citet{piche:08:timse}, is recalled in
  Section~\ref{sec:otherstrat:timse}.}  for the process~$\one_{\xi >
  \thres}$.

\begin{remark}
  The conclusions of Proposition~\ref{prop:ub} still hold in the
  general case when~$\xi$ is not assumed to be a Gaussian process,
  provided that the posterior median~$\bar{\xi}_n$ is substituted to
  posterior the mean~$\xihat_n$.
\end{remark}

\def \hh  {\rule[-1.5ex]{0pt}{4.5ex}}
\def \hhh {\rule[-2.0ex]{0pt}{6.0ex}}

\begingroup

\setbox1=\hbox{\scriptsize
  $\, \bigl( \int \sqrt{\vphantom{|}\tau_{n+1}}\, \dPX \bigr)^2 \,$}
\setbox2=\hbox{\scriptsize
  $\, \bigl( \int \sqrt{\vphantom{|}\nu_{n+1}}\, \dPX \bigr)^2 \,$}
\setbox3=\hbox{\scriptsize
  $\, \int \tau_{n+1} \, \dPX\,$}
\setbox4=\hbox{\scriptsize
  $\, \int \nu_{n+1} \, \dPX\,$}

\wd2=\wd1
\wd3=\wd1
\wd4=\wd1

\begin{table}[htbp]
  \caption{Expressions of four SUR-type criteria.}
  \label{tab:expr-sur-crit}
  \centering \scriptsize
  \begin{tabular}{|l|l|} \hline \hh
    SUR-type sampling criterion     
    & How it is obtained
    \\ \hline \hhh
    $\JSUR{1,n}(x) = \EE_n\Bigl( \box1 \Bigm| X_{n+1} = x 
    \Bigr)$
    & Prop.~\ref{prop:ub} with $\ahat_n = \int \one_{\xihat_n > \thres}\, \dPX$
    \\ \hhh
    $\JSUR{2,n}(x) = \EE_n\Bigl( \box2 \Bigm| X_{n+1} = x 
    \Bigr)$
    & Prop.~\ref{prop:ub} with $\ahat_n = \EE_n\left( \alpha \right)$     
    \\ \hhh
    $\JSUR{3,n}(x) = \EE_n\Bigl( \box3 \Bigm| X_{n+1} = x 
    \Bigr)$
    & Eq.~\eqref{equ:sur-ub-2} with $\ahat_n = \int \one_{\xihat_n > \thres}\, \dPX$
    \\ \hhh
    $\JSUR{4,n}(x) = \EE_n\Bigl( \box4 \Bigm| X_{n+1} = x \Bigr)$    
    & Eq.~\eqref{equ:sur-ub-2} with $\ahat_n = \EE_n\left( \alpha \right)$ 
    \\ \hline
  \end{tabular}
\end{table}

\endgroup

\subsection{Discretizations}
\label{sec:SUR-discrete}

In this section, we proceed with the necessary integral discretizations
of the SUR criteria to make them suitable for numerical evaluation and
implementation on computers.  Assume that $n$ steps of the algorithm
have already been performed and consider, for instance, the 
criterion 
\begin{equation}
  \label{equ:JSUR3:form1}
  \JSUR{3,n}(x) = \EE_n
  \Bigl( 
  \smallint \tau_{n+1}(y)\, \PX(\dy) \Bigm| X_{n+1} = x 
  \Bigr) .
\end{equation}
Remember that, for each $y\in \XX$, the probability of
misclassification~$\tau_{n+1}(y)$ is $\FF_{n+1}$-measurable and,
therefore, is a function of $\Ical_{n+1} = \left( \Ical_n,
  X_{n+1}, Z_{n+1} \right)$. Since $\Ical_n$ is known at this point,
we introduce the notation $v_{n+1}(y;X_{n+1},Z_{n+1}) =
\tau_{n+1}(y)$ to emphasize the fact that,  when a new
evaluation point must be chosen at step $(n+1)$,
$\tau_{n+1}(y)$ depends on the choice of $X_{n+1}$ and the random outcome $Z_{n+1}$. Let us
further denote by $\Q_{n,x}$ the probability distribution of~$\xi(x)$
under~$\P_n$. Then, \eqref{equ:JSUR3:form1} can be rewritten
as
\begin{equation*}
  \JSUR{3,n}(x) = \iint_{\RR \times \XX}
   v_{n+1}(y;x,z)\; \Q_{n,x}(\dz)\, \PX(\dy)\,,
\end{equation*}
and the corresponding strategy is:
\begin{equation}
  \label{equ:JSUR3:form2}
  X_{n+1} \;=\;\argmin_{x\in\XX} \iint_{\RR \times \XX}
  v_{n+1}(y;x,z)\; \Q_{n,x}(\dz)\, \PX(\dy)\,.
\end{equation}
Given~$\Ical_n$ and a triple $(x,y,z)$, $v_{n+1}(y;x,z)$ can be computed
efficiently using the equations provided in
Sections~\ref{sec:gauss-proc-priors} and~\ref{sec:estimators}.

At this point, we
need to address: 1) the computation of the integral on~$\XX$ with
respect to~$\PX$; 2) the computation of the integral on~$\RR$ with
respect to~$\Q_{n,x}$; 3) the minimization of the resulting criterion
with respect to~$x \in \XX$.

To solve the first problem, we draw an i.i.d. sequence $Y_1, \ldots,
Y_m \sim \PX$ and use the Monte Carlo approximation:
\begin{equation*}
  \int_{\XX} v_{n+1}(y;x,z)\; \PX(\dy)
  \;\approx\;
  \frac{1}{m} \sum_{j=1}^m v_{n+1}(Y_j;x,z) .
\end{equation*}
An increasing sample size $n \mapsto m_n$ should be used to build a
convergent algorithm for the estimation of~$\alpha$ (possibly with a
different sequence $Y_{n,1}, \ldots, Y_{n,m_n}$ at each step). In this
paper we adopt a different approach instead, which is to take a fixed
sample size~$m > 0$ and keep the same sample $Y_1, \ldots, Y_m$
throughout the iterations.
Equivalently, it means that we choose to work from the start on a
discretized version of the problem: we replace $\PX$ by the empirical
distribution $\hat{\P}_{\XX,n} = \frac{1}{m} \sum_{j=1}^m
\delta_{Y_j}$, and our goal is now to \emph{estimate the Monte Carlo estimator
  $\alpha_m = \int \one_{\xi>u} \ddiff \hat{\P}_{\XX,n} = \frac{1}{m}
  \sum_{j=1}^m \one_{\xi(Y_j)>u}$}, using either the posterior mean
$\EE_n\left( \alpha_m \right) = \frac{1}{m} \sum_j p_n(Y_j)$ or the
plug-in estimate $\frac{1}{m} \sum_j \one_{\xihat(Y_j;\Xv_n) > u
}$. This kind of approach has be coined \emph{meta-estimation} by
\citet{arnaud:2010:jds}: the objective is to estimate the value
of a precise Monte Carlo estimator of $\alpha(f)$ ($m$ being large),
using prior information on~$f$ to alleviate the
computational burden of running $m$ times the computer code~$f$. This
point of view also underlies the work in structural reliability of
\citet{hurtado:2004, hurtado:2007}, \citet{deheeger:2007:svm},
\citet{deheeger:2008}, and more recently \citet{echard:2010, echard:2010:b}.

The new point of view also suggests a natural solution for the third
problem, which is to replace the continuous search for a
minimizer~$x\in \XX$ by a discrete search over the set $\XX_m :=
\left\{ Y_1, \ldots, Y_m \right\}$. This is obviously sub-optimal,
even in the meta-estimation framework introduced above, since picking
$x \in \XX\setminus\XX_m$ can sometimes bring more information
about~$\xi(Y_1),\ldots, \xi(Y_m)$ than the best possible choice
in~$\XX_m$. Global optimization algorithms may of course be used to
tackle directly the continuous search problem: for instance,
\cite{ranjan:2008} use a combination of a genetic algorithm and
local search technique, \cite{bichon:2008} use the DIRECT algorithm
and \cite{piche:08:timse} use a covariance-matrix-adaptation evolution strategy. In this
paper we will stick to the discrete search approach, since it is much
simpler to implement (we shall present in~Section~\ref{sec:SUR:impl}  a
method to handle the case of large~$m$) and provides
satisfactory results (see Section~\ref{sec:numerical}).

Finally, remark that the second problem boils down to the computation
of a one-dimensional integral with respect to Lebesgue's measure. Indeed,
since~$\xi$ is a Gaussian process, $\Q_{n,x}$ is a Gaussian
probability distribution with mean~$\xihat_n(x)$ and
variance~$\sigma_n^2(x)$ as explained in
Section~\ref{sec:gauss-proc-priors}. The integral can be computed  using a
standard Gauss-Hermite quadrature with~$Q$ points \citep[see,
e.g.,][Chapter~4]{press:1992:nrc} :
\begin{equation*}
  \int v_{n+1}(y;x,z)\, \Q_{n,x}(\dz)
  \;\approx\;
  \frac{1}{\sqrt{\pi}}
  \sum_{q=1}^Q w_q\, v_{n+1}(y;x,\xihat_n(x)+ \sigma_n(x) u_q\sqrt{2} ) \,,
\end{equation*}
where $u_1, \ldots, u_Q$ denote the quadrature points and $w_1,
\ldots, w_Q$ the corresponding weights. Note that this is equivalent
to replacing under~$\P_n$ the random variable~$\xi(x)$ by a quantized
random variable with probability distribution $\sum_{q=1}^Q w'_q
\delta_{z_{n+1,q}(x)}$, where $w'_q = w_q/\sqrt{\pi}$ and $z_{n+1,q}(x) =
\xihat_n(x)+ \sigma_n(x) u_q\sqrt{2}$.

Taking all three discretizations into account, the proposed strategy is:
\begin{equation}
  \label{equ:discr-strat}
  X_{n+1} \;=\; \argmin_{1 \le k  \le m}\,
  \sum_{j=1}^m\, \sum_{q=1}^Q\, w'_q\,
  v_{n+1}\left( Y_j;\, Y_k, z_{n+1,q}(Y_k) \right).
\end{equation} 
\pagebreak[0]

\subsection{Implementation}
\label{sec:SUR:impl}

This section gives implementation guidelines for the SUR strategies
described in Section~\ref{sec:SUR}. As said in
Section~\ref{sec:SUR-discrete}, the strategy~(\ref{equ:discr-strat})
can, in principle, be translated directly into a computer program. In
practice however, we feel that there is still room for different
implementations. In particular, it is important to keep the
computational complexity of the strategies at a reasonable level. We
shall explain in this section some simplifications we have made to
achieve this goal.

A straight implementation of~(\ref{equ:discr-strat}) for the choice of
an additional evaluation point is described in Table~\ref{tab:SUR}. This
procedure is meant to be called iteratively in a sequential algorithm,
such as that described for instance in Table~\ref{tab:seqalgo}. Note
that the only parameter to be specified in the SUR
strategy~\eqref{equ:discr-strat} is~$Q$, which tunes the precision of
the approximation of the integral on $\RR$ with respect to
$\mathsf{Q}_{n,x}$. In our numerical experiments, it was observed that
taking $Q=12$ achieves a good compromise between precision and numerical complexity.

\begin{table}[htbp]
  \caption{Procedure to select a new evaluation point $X_{n+1}\in\XX$
    using a SUR strategy}
  \label{tab:SUR}
  \noindent\ignorespaces%
  \rule{\textwidth}{.2pt}%
  \par
  Require computer representations of
  \begin{enumerate}[a)]
  \item a set $\Ical_n = \{(X_1, f(X_1)),\ldots,(X_n,f(X_n))\}$ of evaluation results;
  \item a Gaussian process prior $\xi$ with a (possibly unknown linear
    parametric) mean function and a covariance function $k_{\theta}$,
    with parameter $\theta$;
  \item a (pseudo-)random sample $\XX_m = \{Y_1,\ldots, Y_m\}$ of size
    $m$ drawn from the
    distribution $\P_{\XX}$;
  \item quadrature points $u_1,\ldots,u_Q$ and corresponding weights
    $w'_1,\ldots, w'_Q$;
  \item a threshold $u$.
  \end{enumerate}
  \medskip

  \begin{description}
  \item[1.] compute the kriging approximation $\hat f_n$ and kriging
    variance $\sigma_n^2$ on $\XX_m$ from $\Ical_n$
  \item[2.] for each candidate point $Y_j$, $j \in \{1,\ldots,m\}$,
    \begin{description}
    \item[2.1] for each point $Y_k$, $k\in \{1,\ldots,m\}$, compute
      the kriging weights $\lambda_{i}(Y_k;\{\Xv_n, Y_j\})$,
      $i\in\{1,\ldots,(n+1)\}$, and the kriging variances
      $\sigma^2(Y_k;\{\Xv_n, Y_j\})$
    \item[2.2] compute $z_{n+1,q}(Y_j) = \hat f_n(Y_j) + \sigma_n(Y_j)
      u_q \sqrt{2}$, for $q=1,\ldots, Q$
    \item[2.3] for each $z_{n+1,q}(Y_j)$, $q\in\{1,\ldots, Q\}$,

      \begin{description}
      \item[2.3.1] compute the kriging approximation $\tilde
        f_{n+1,j,q}$ on $\XX_m$ from $\Ical_n \cup \left(Y_j, f(Y_j) =
          z_{n+1,q}(Y_j)\right)$, using the weights
        $\lambda_{i}(Y_k;\{\Xv_n, Y_j\})$, $i=1,\ldots,(n+1)$,
        $k=1,\ldots,m$, obtained at Step 2.1.
      \item[2.3.2] for each $k\in \{1,\ldots,m\}$, compute
        $v_{n+1}\left( Y_k;\, Y_j, z_{n+1,q}(Y_j) \right)$, using $u$,
        $\tilde f_{n+1,j, q}$ obtained in 2.3.1, and
        $\sigma^2(Y_k;\{\Xv_n, Y_j\})$ obtained in 2.1
      \end{description}
      
    \item[2.4] compute $J_n(Y_j) = \sum_{k=1}^m\, \sum_{q=1}^Q\, w'_q\,
      v_{n+1}\left( Y_k;\, Y_j, z_{n+1,q}(Y_j) \right)$.
    \end{description}
    
  \item[3.] find $j^{\star} = \argmin_{j} J_n(Y_j)$ and set $X_{n+1} =
    Y_{j^{\star}}$
  \end{description}
  \noindent\ignorespaces%
  \rule{\textwidth}{.2pt}%
\end{table}

\begin{table}[htpb]
  \caption{Sequential estimation of a probability of failure}
  \label{tab:seqalgo}
  \noindent\ignorespaces%
  \rule{\textwidth}{.2pt}%
  \par
  \begin{description}

  \item[1.] Construct an initial design of size $n_0<N$ and evaluate $f$
    at the points of the initial design.
  \item[2.] Choose a Gaussian process $\xi$ (in practice, this amounts
    to choosing a parametric form for the mean of $\xi$ and a parametric
    covariance function $k_{\theta}$)
  \item[3.] Generate a Monte Carlo sample $\XX_m = \{Y_1,\ldots, Y_m\}$ of size $m$ from $\P_{\XX}$
  \item[4.] While the evaluation budget $N$ is not exhausted,
    \begin{description}
    \item[4.1] optional step: estimate the parameters of the covariance
      function (case of a plug-in approach);
    \item[4.2] select a new evaluation point, using past evaluation
      results, the prior $\xi$ and $\XX_m$;
    \item[4.3] perform the new evaluation.
    \end{description}
  \item[5.] Estimate  the probability of failure
    obtained from the $N$ evaluations of $f$ (for instance, by using $\EE_N\left( \alpha_m \right) = \frac{1}{m} \sum_j p_N(Y_j)$).
  \end{description}
  \noindent\ignorespaces%
  \rule{\textwidth}{.2pt}%
\end{table}

To assess the complexity of a SUR sampling strategy, recall that kriging
takes ${\rm O}(mn^2)$ operations to predict the value of $f$ at $m$
locations from $n$ evaluation results of $f$ (we suppose that $m>n$ and
no approximation is carried out). In the procedure to select an
evaluation, a first kriging prediction is performed at Step 1 and then,
$m$ different predictions have to performed at step 2.1.  This cost
becomes rapidly burdensome for large values of $n$ and $m$, and we must
further simplify~(\ref{equ:discr-strat}) to be able to work on
applications where $m$ must be large. A natural idea to alleviate the
computational cost of the strategy is to avoid dealing with candidate
points that have a very low probability of misclassification, since they
are probably far from the frontier of the domain of failure. It is also
likely that those points with a low probability of misclassification
will have a very small contribution in the variance of the error of
estimation $\ahat_n - \alpha_m$.

Therefore, the idea is to rewrite the sampling strategy described
by~(\ref{equ:discr-strat}), in such a way that the first summation (over
$m$) and the search set for the minimizer is restricted to a subset of
points $Y_j$ corresponding to the $m_0$ largest values of
$\tau_n(Y_j)$. The corresponding algorithm is not described here for the
sake of brevity but can easily be adapted from that of
Table~\ref{tab:SUR}. Sections~\ref{sec:num:structural}
and~\ref{sec:num:average} will show that this \emph{pruning} scheme has
almost no consequence on the performances of the SUR strategies, even
when one considers small values for $m_0$ (for instance $m_0=200$).

\section{Other strategies proposed in the literature}
\label{sec:otherstrat}

\subsection{Estimation of a probability of failure and closely related
  objectives}
\label{sec:related-to-estim-prob-fail}

Given a real function $f$ defined over $\XX\subseteq \RR^{d}$, and a
threshold $\thres \in \RR$, consider the following possible goals:
\begin{enumerate}
\item estimate a region $\Gamma \subset \XX$ of the form $\Gamma =
  \{ x \in \XX \bmid f(x) > u \}$;
\item estimate the level set $\partial \Gamma = \{ x \in
  \XX \bmid f(x) = u \}$;
\item estimate $f$ precisely in a neighborhood of $\partial \Gamma$;
\item estimate the probability of failure $\alpha = \PX(\Gamma)$ for
  a given probability measure~$\PX$.
\end{enumerate}
These different goals are, in fact, closely related: indeed, they all
require, more or less explicitly, to select sampling points in order
to get a fine knowledge of the function~$f$ in a neighborhood of the
level set $\partial \Gamma$ (the location of which is unknown before
the first evaluation).  Any strategy proposed for one of the first
three objectives is therefore expected to perform reasonably well on
the fourth one, which is the topic of this paper.

Several strategies recently introduced in the literature are presented
in Sections~\ref{sec:otherstrat:timse} and~\ref{sec:otherstrat:pointwise},
and will be compared numerically to the SUR strategy in
Section~\ref{sec:numerical}. Each of these strategies has been
initially proposed by its authors to address one or several of the
above objectives, but they will only be discussed in this paper from
the point of view of their performance on the fourth one. Of course, a
comparison focused on any other objective would probably be based on
different performance metrics, and thus could yield a different
performance ranking of the strategies.

\subsection{The targeted IMSE criterion}
\label{sec:otherstrat:timse}

The \emph{targeted IMSE} proposed in \citet{piche:08:timse} is a
modification of the IMSE (Integrated Mean Square Error) sampling
criterion \citep{sac89}. While the IMSE sampling criterion computes the
average of the kriging variance (over a compact domain $\XX$) in order
to achieve a space-filling design, the targeted IMSE computes a weighted
average of the kriging variance for a better exploration of the regions
near the frontier of the domain of failure, as in \cite{oakley:2004:perc}. The idea is to put a large
weight in regions where the kriging prediction is close to the
threshold~$\thres$, and a small one otherwise. Given $\Ical_n$, the
targeted IMSE sampling criterion, hereafter abbreviated as tIMSE, can be
written as
\begin{align}  
  \JTIMSE{n}(x) 
  & \;=\; \EE_n \left(\,
    \int_\XX \bigl( \xi - \xihat_{n+1} \bigr)^2\, W_n\, \dPX
    \Bigm| X_{n+1}  = x
  \right)
  \label{eq:tIMSE1} \\
  & \;=\; \int_\XX \sigma^2\left(y;X_1,\ldots,X_n,x\right)\,
  W_n(y)\, \PX(\dy),
  \label{eq:tIMSE2}  
\end{align}
where $W_n$ is a weight function based on $\Ical_n$. The weight function
suggested by \citet{piche:08:timse} is
\def\sn{s_n(x)}
\begin{equation}
  W_n(x) \;=\; \frac{1}{\sn\, \sqrt{2\pi}}\;
  \exp\left( - \frac{1}{2}\, \biggl( \frac{\xihat_n(x) - \thres}{\sn} \biggr)^2 \right),
\end{equation}
where $s_n^2(x) = \se ^2 + \sigma_n^2\left( {x} \right)$. Note that
$W_n(x)$ is large when $\xihat_n(x) \approx \thres$ and $\sigma_n^2(x)
\approx 0$, i.e., when the function is known to be close
to~$\thres$. 

The tIMSE criterion operates a trade-off between global uncertainty
reduction (high kriging variance $\sigma_n^2$) and exploration of target
regions (high weight function~$W_n$). The weight function depends on a
parameter $\se > 0$, which allows to tune the width of the ``window of
interest'' around the threshold. For large values of~$\se$, $\JTIMSE{}$
behaves approximately like the IMSE sampling criterion. The choice of an
appropriate value for~$\se$, when the goal is to estimate a probability
of failure, will be discussed on the basis of numerical experiments in
Section~\ref{sec:num:average}.

The tIMSE strategy requires a computation of the expectation with respect
to~$\xi(x)$ in~\eqref{eq:tIMSE1}, which can be done analytically,
yielding~\eqref{eq:tIMSE2}. The computation of the integral with respect
to~$\PX$ on~$\XX$ can be carried out with a Monte Carlo approach, as
explained in Section~\ref{sec:SUR-discrete}. Finally, the optimization of
the criterion is replaced by a discrete search in our implementation.

\subsection{Criteria based on the marginal distributions}
\label{sec:otherstrat:pointwise}

Other sampling criteria proposed by \citet{ranjan:2008},
\citet{bichon:2008} and \citet{echard:2010, echard:2010:b} are briefly
reviewed in this section\footnote{Note that the paper of \citet{ranjan:2008} is
  the only one in this category that does not address the problem of
  estimating a probability of failure (i.e., Objective~4 of
  Section~\ref{sec:related-to-estim-prob-fail}).}. A common feature
of these three criteria is that, unlike the SUR and
tIMSE criteria discussed so far, they only depend on the \emph{marginal
  posterior distribution} at the considered candidate point $x \in \XX$,
which is a Gaussian $\Ncal \bigl( \xihat_n(x), \sigma_n^{2}(x) \bigr)$
distribution. As a consequence, they are of course much cheaper to
compute than integral criteria like SUR and tIMSE.

A natural idea, in order to sequentially improve the estimation of the
probability of failure, is to visit the point~$x \in \XX$ where the event $\{\xi(x) \geq
u\}$ is the most uncertain. This idea, which has been explored by
\citet*{echard:2010, echard:2010:b}, corresponds formally to the sampling criterion
\begin{equation}
  \label{eq:crit-echard}
  \JECH{n}(x) \;=\; \tau_n(x) 
  \;=\; 1 - \normCDF\left( 
    \frac{\bigl| u-\xihat_n(x) \bigr|}{\sigma_n(x)} 
  \right) \,.
\end{equation}
As in the case of the tIMSE criterion and also, less explicitly, in
SUR criteria, a trade-off is realized between global uncertainty
reduction (choosing points with a high~$\sigma_n^2(x)$) and
exploration of the neighborhood of the estimated contour (where
$\bigl| u - \xihat_n(x) \bigr|$ is small).

The same leading principle motivates the criteria proposed by
\citet{ranjan:2008} and \citet{bichon:2008}, which can be seen as
special cases of the following sampling criterion:
\begin{equation}
  \label{RanjBichLikeCriteria}
  \JRB{n}(x) \;:=\;
  \EE_n \left( \max\left( 0, 
      \epsilon(x)^{\delta} - \left|u-\xi(x)\right|^{\delta} 
    \right) \right),
\end{equation}
where $\epsilon(x) = \kappa\, \sigma_n(x)$, $\kappa, \delta > 0$. The
following proposition provides some insights into this sampling criterion:
\begin{proposition}
  \label{prop:crit-RB}
  Define $G_{\kappa,\delta}: \left] 0, 1\right[ \to \Rplus$ by
    \begin{equation*}
      G_{\kappa,\delta}(p) \;:=\; 
      \EE\left( 
        \max\Bigl( 
        0, \kappa^\delta - \bigl| \normCDF^{-1}(p) + U \bigr| 
        \Bigr)
      \right),
    \end{equation*}
    where $U$ is a Gaussian $\Ncal(0,1)$ random variable. Let~$\normPDF$
    and~$\normCDF$ denote respectively the probability density function
    and the cumulative distribution function of~$U$.
    \begin{enumerate}[a) ]
      % 
      % G has the property that we expect !
      % 
    \item\label{prop:G:sym} $G_{\kappa,\delta}(p) =
      G_{\kappa,\delta}(1-p)$ for all $p \in\, ]0,1[$.
    \item $G_{\kappa,\delta}$ is strictly increasing on $\left] 0, 1/2
      \right]$ and vanishes at~$0$.  Therefore, $G_{\kappa,\delta}$ is
      also strictly decreasing on $\left[ 1/2, 1 \right[$, vanishes
      at~$1$, and has a unique maximum at~$p = 1/2$.
      % 
      % better expression for \JRB
      % 
    \item\label{prop:JRB-func-G}
      Criterion~(\ref{RanjBichLikeCriteria}) can be rewritten as
      \begin{equation}
        \label{eq:RBsamplified}
        \JRB{n}(x) = \sigma_n(x)^\delta\, G_{\kappa,\delta}\bigl( p_n(x) \bigr).
      \end{equation}
      % 
      % closed-form expression for Bichon
      % 
    \item\label{prop:cf-bichon} $G_{\kappa,1}$ has the following
      closed-form expression:
      \begin{equation}
        \begin{aligned}
          G_{\kappa,1}(p) \;=
          &\;\ \kappa\, \bigl( \normCDF(t^+) - \normCDF(t^-) \bigr)\\ 
          &\; -\, t\, \bigl( 2 \normCDF(t) - \normCDF(t^+) - \normCDF(t^-) \bigr)\\
          &\; -\, \bigl( 2 \normPDF(t) - \normPDF(t^+) - \normPDF(t^-) \bigr),
        \end{aligned}
        \label{eq:bichon-explicit}
      \end{equation}
      where $t = \normCDF^{-1}(1-p)$, $t^+ = t + \kappa$ and $t^- = t -
      \kappa$.
      %
      % closed-form (really?) expression for Ranjan
      % 
    \item\label{prop:cf-ranjan} $G_{\kappa,2}$ has the following
      closed-form expression:
      \begin{equation}
        \begin{aligned}
          G_{\kappa,2}(p) \;=
          &\; \ \bigl( \kappa^2 - 1 - t^2 \bigr)\,
                \bigl( \normCDF(t^+) - \normCDF(t^-) \bigr)\\ 
          &\; -\, 2 t\, \bigl( \normPDF(t^+) - \normPDF(t^-) \bigr) \\
          &\; +\, t^+ \normPDF(t^+) - t^- \normPDF(t^-) ,
        \end{aligned}
        \label{eq:ranjan:explicit}
      \end{equation}
      with the same notations.
    \end{enumerate}
\end{proposition}

\noindent It follows from~\ref{prop:G:sym}) and~\ref{prop:JRB-func-G}) 
that $\JRB{n}(x)$ can also be seen
as a function of the kriging variance~$\sigma_n^2(x)$ and the
probability of misclassification~$\tau_n(x) = \min\left( p_n(x), 1 -
  p_n(x) \right)$. Note that, in the computation
of~$G_{\kappa,\delta}\bigl( p_n(x) \bigr)$, the quantity denoted
by~$t$ in~\eqref{eq:bichon-explicit} and~\eqref{eq:ranjan:explicit} is
equal to $\bigl( u - \xihat_n(x) \bigr) / \sigma_n(x)$, i.e., equal to
the normalized distance between the predicted value and the threshold.

\citeauthor{bichon:2008}'s \textit{expected feasibility} function
corresponds to~\eqref{eq:RBsamplified} with~$\delta=1$, and can be
computed efficiently using~\eqref{eq:bichon-explicit}. Similarly,
\citeauthor{ranjan:2008}'s \textit{expected
  improvement}\footnote{Despite its name and some similarity between
  the formulas, this criterion should not be confused with the
  well-known EI criterion in the field of optimization \citep{Mockus,
    Jones}.}  function corresponds to~\eqref{eq:RBsamplified}
with~$\delta = 2$, and can be computed efficiently
using~\eqref{eq:ranjan:explicit}. The proof of
Proposition~\ref{prop:crit-RB} is provided in
Appendix~\ref{sec:comput:rb}.

\begin{remark}
  In the case~$\delta = 1$, our result coincides with the expression
  given by \citet[][Eq.~(17)]{bichon:2008}. In the case $\delta = 2$,
  we have found and corrected a mistake in the computations of
  \citet[][Eq.~(8) and Appendix~B]{ranjan:2008}.
\end{remark}

\section{Numerical experiments}
\label{sec:numerical}

\subsection{A one-dimensional illustration of a SUR strategy}
\label{sec:1D-illustration}

\begin{figure}[p]
  \centering
  \includegraphics[width=0.6\textwidth]{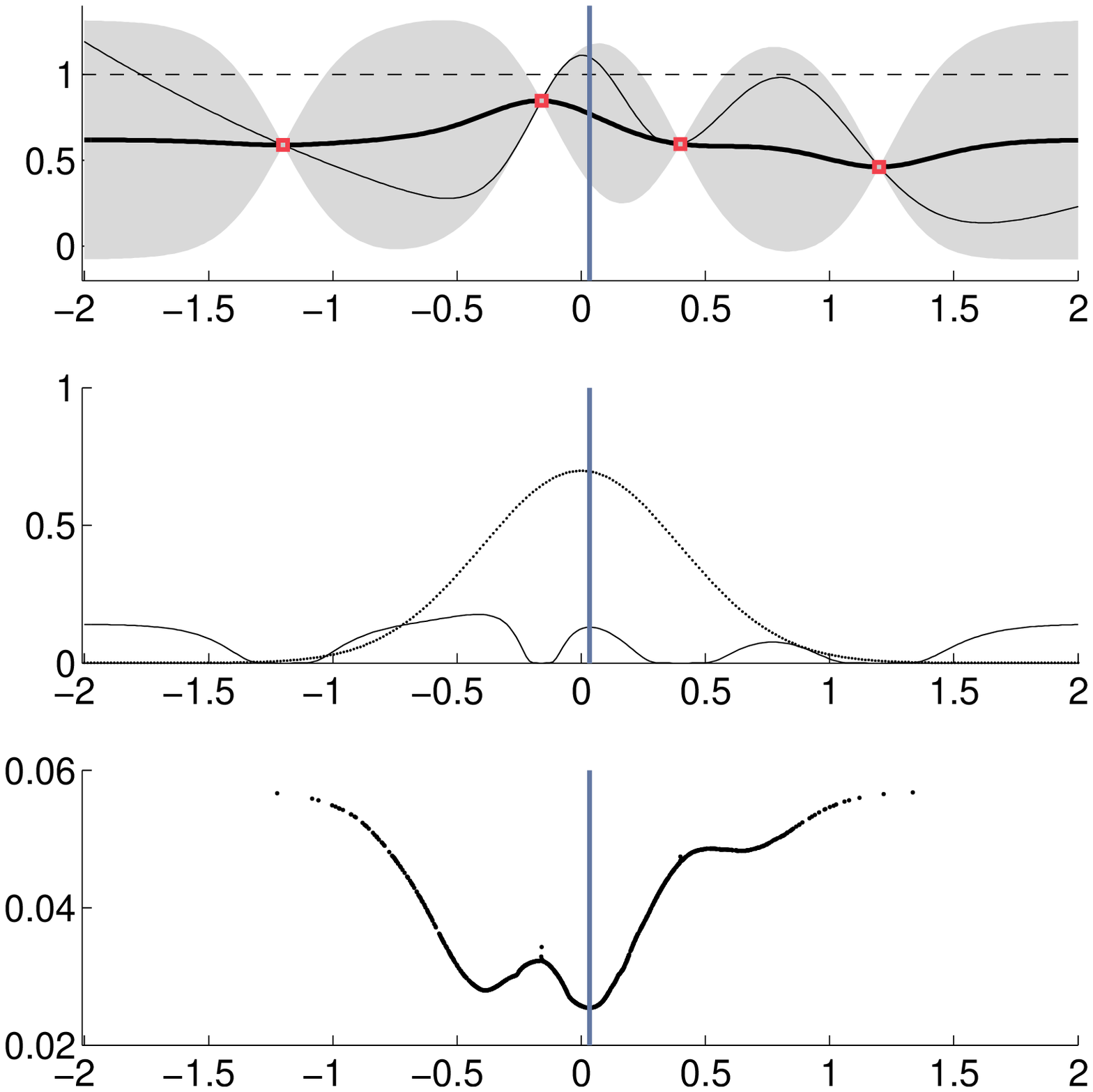}
  \caption{Illustration of a SUR strategy. This figure shows the initial
    design. Top: threshold $u=1$ (horizontal dashed line); function $f$
    (thin line); $n=4$ initial evaluations (squares); kriging
    approximation $f_n$ (thick line); 95\% confidence intervals computed
    from the kriging variance (shaded area).  Middle: probability of
    excursion (solid line); probability density of $\P_\XX$ (dotted
    line). Bottom: graph of $\JSUR{1,n=4}(Y_i)$, $i=1,\ldots,m=1500$, the minimum
    of which indicates where the next evaluation of $f$ should be done
    (i.e., near the origin).}
  \label{fig:1D-1}
\end{figure}

The objective of this section is to show the progress of a SUR strategy
in a simple one-dimensional case.  We wish to
estimate~$\alpha=\P_{\XX}\{f>1\}$, where $f:\XX=\RR\to\RR$ is such that
$\forall x\in\RR$,
$$
f(x)= (0.4 x-0.3)^2 + \exp\left(-11.534\,\abs{x}^{1.95}\right)+\exp(-5(x-0.8)^2)\,,
$$
and where $\XX$ is endowed with the probability distribution
$\P_{\XX}=\mathcal{N}(0,\sigma_{\XX}^2)$, $\sigma_{\XX}=0.4$, as
depicted in Figure~\ref{fig:1D-1}. We know in advance that
$\alpha\approx 0.2$. Thus, a Monte Carlo sample of size $m=1500$ will
give a good estimate of $\alpha$.

\begin{figure}[p]
  \centering
  \includegraphics[width=0.6\textwidth]{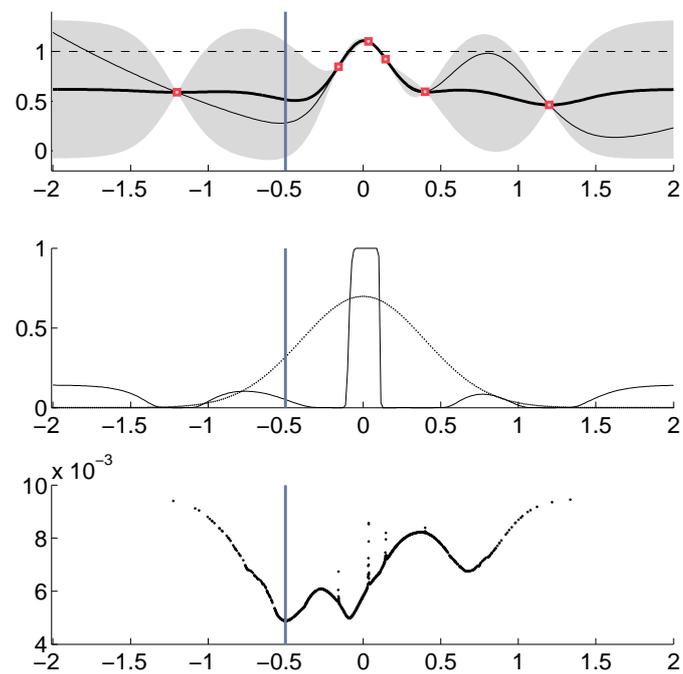}
  \caption{Illustration of a SUR strategy (see also
    Figures~\ref{fig:1D-1} and \ref{fig:1D-3}). This figure shows the
    progress of the SUR strategy after two iterations---a total of $n=6$ evaluations
    (squares) have been performed. The next evaluation point will be
    approximately at $x=-0.5$}.
  \label{fig:1D-2}
\end{figure}

\begin{figure}[htbp]
  \centering
  \includegraphics[width=0.6\textwidth]{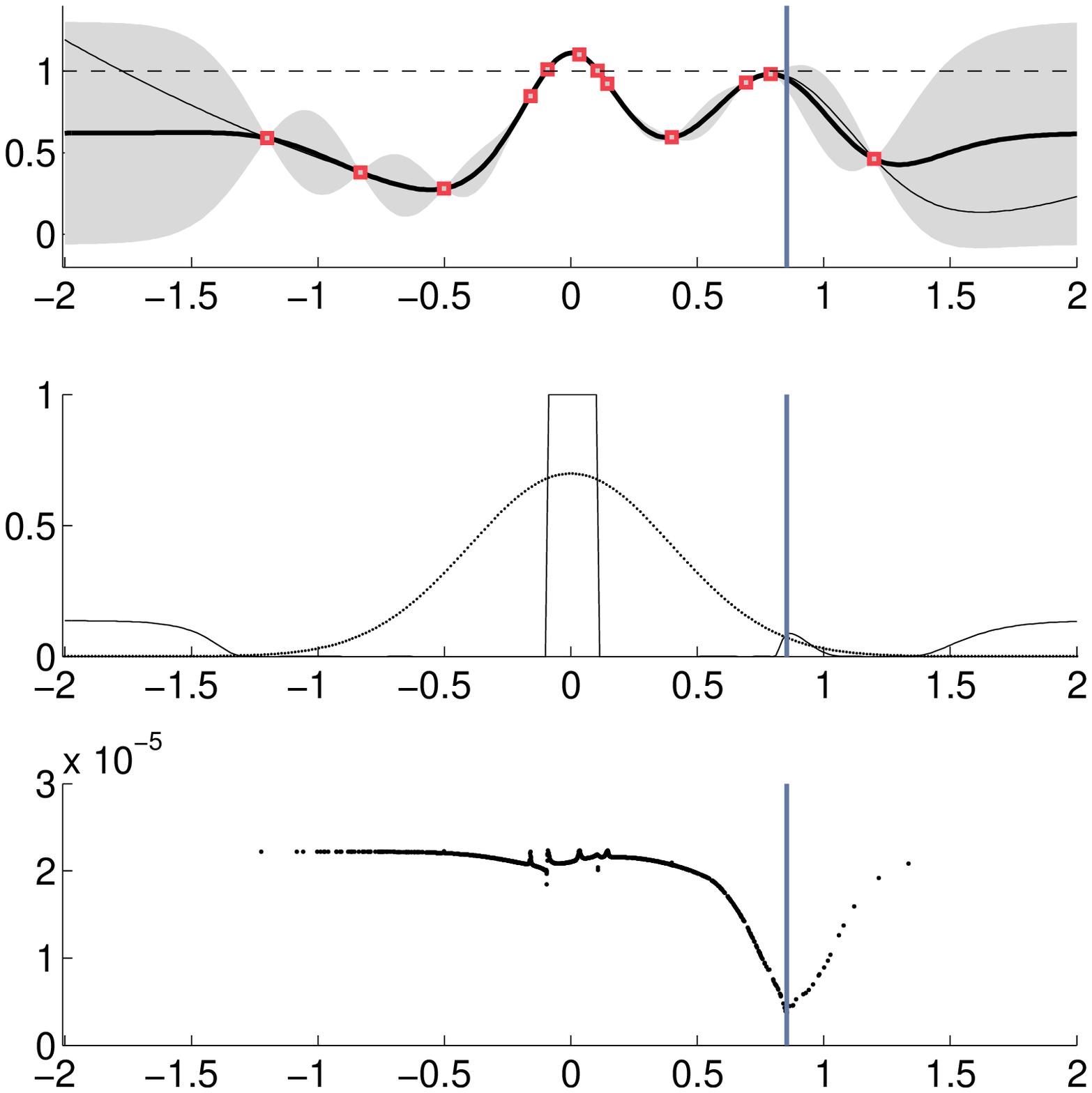}
  \caption{Illustration of a SUR strategy (see also
    Figures~\ref{fig:1D-1} and \ref{fig:1D-2}). This figure shows the
    progress of the SUR strategy after eight iterations---a total of
    $n=12$ evaluations (squares) have been performed. At this stage, the
    probability of excursion $p_n$ almost equals $0$ or $1$ in the
    region where the density of $\P_{\XX}$ is high.}
  \label{fig:1D-3}
\end{figure}

In this illustration, $\xi$ is a Gaussian process with constant but
unknown mean and a Mat\'ern covariance function, whose parameters are
kept \emph{fixed}, for the sake of simplicity.  Figure~\ref{fig:1D-1}
shows an initial design of four points and the sampling criterion
$\JSUR{1,n=4}$. Notice that the sampling criterion is only computed at
the points of the Monte Carlo sample. Figures~\ref{fig:1D-2} and
\ref{fig:1D-3} show the progress of the SUR strategy after a few
iterations. Observe that the unknown function $f$ is sampled so that the
probability of excursion $p_n$ almost equals zero or one in the region
where the density of $\P_{\XX}$ is high.


\subsection{An example in structural reliability}
\label{sec:num:structural}

\begin{figure}[ht!]
 \centering
  \includegraphics[width = 6cm, height = 5.5 cm]{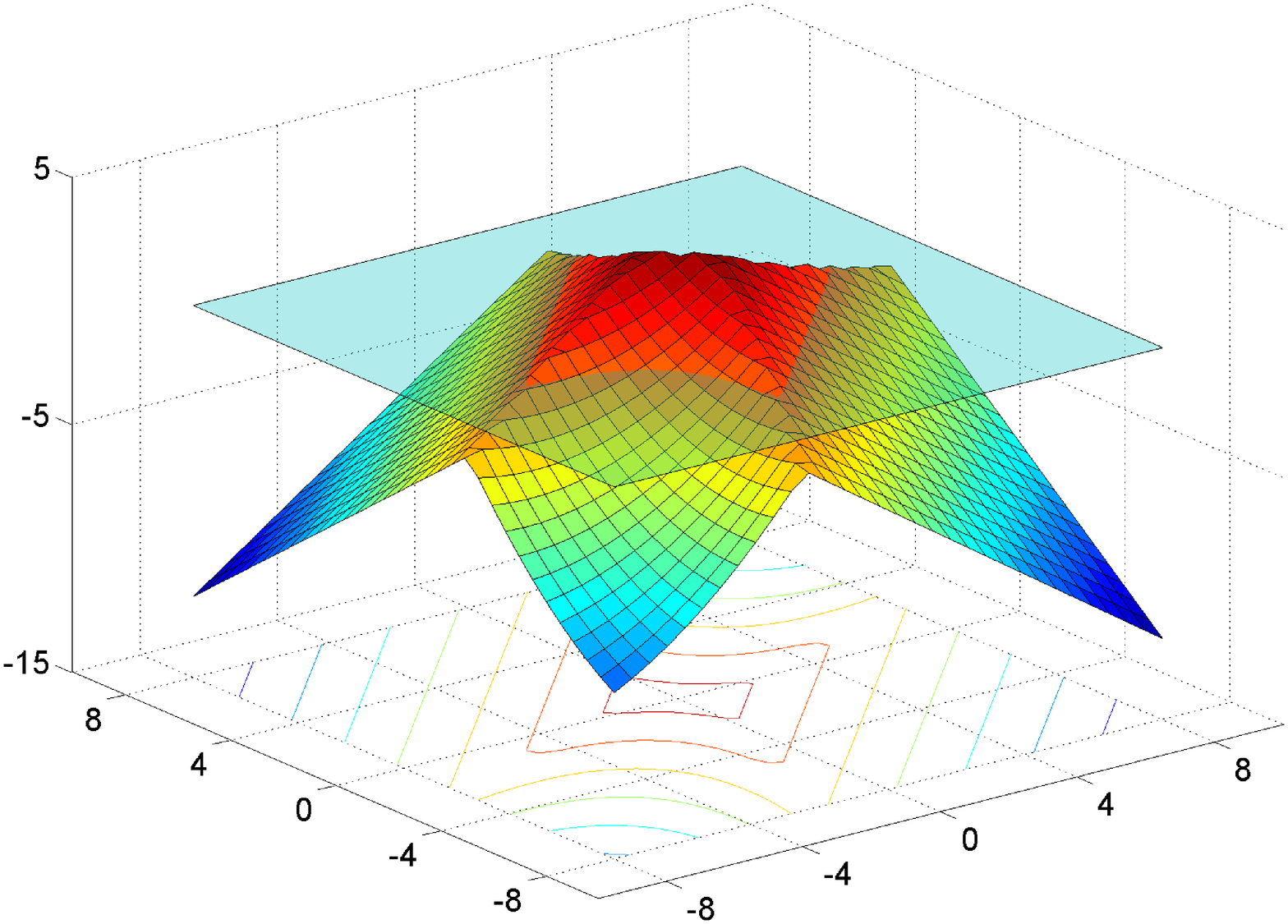}
  \psfrag{fff}{$f=0$}
  \includegraphics[width = 6cm, height = 5.5 cm]{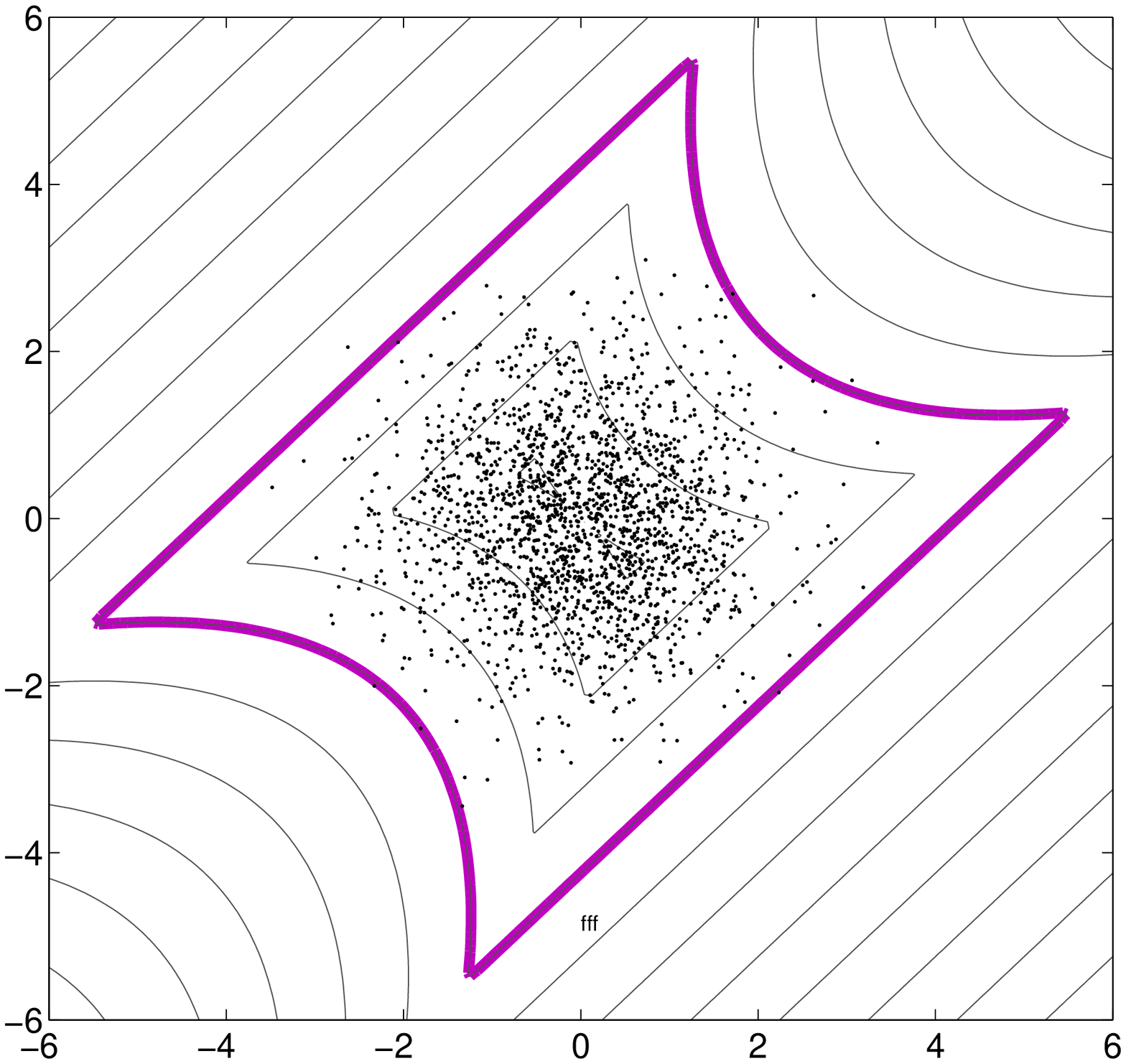}
 \caption{Left: mesh plot of the performance function $f$ corresponding to the
   four-branch series system; a
   failure happens when $f$ is below the transparent plane;  
   Right: contour plot of $f$; limit state $f=0$ (thick
   line); sample of size $m=3\times10^{3}$  from $\P_{\XX}$~(dots).}
\label{funcshow}
\end{figure}

In this section, we evaluate all criteria discussed in
Section~\ref{sec:SUR} and Section~\ref{sec:otherstrat} through a
classical benchmark example in structural reliability \citep[see,
e.g.,][]{ borri:1997, waarts:2000, schueremans:2001, deheeger:2008}.
\cite{echard:2010, echard:2010:b} used this benchmark to make  a comparison among several methods proposed in
\cite{schueremans:2005:benefit}, some of which are based on the
construction of a response surface. The objective of the benchmark is to
estimate the probability of failure of a so-called \emph{four-branch
  series system}.  A failure happens when the system is working under the
threshold $\thres=0$. The performance function $f$ for this system is
defined as
\begin{equation}
f: (x_1,x_2)\in \RR^2 \mapsto f(x_1,x_2) = \min \left\{
\begin{array}{ll}
3+0.1(x_1-x_2)^{2}-(x_1+x_2)/ \sqrt{2}; \nonumber \\
3+0.1(x_1-x_2)^{2}+(x_1+x_2)/ \sqrt{2}; \nonumber \\
(x_1-x_2)+6/ \sqrt{2}; \nonumber \\
(x_2-x_1)+6/ \sqrt{2}
\end{array} \right \}\,.
\end{equation}   
The uncertain input factors are supposed to be independent and have
standard normal distribution.  Figure~\ref{funcshow} shows the
performance function, the failure domain and the input
distribution. Observe that $f$ has a first-derivative discontinuity
along four straight lines originating from the point $(0,0)$.

For each sequential method, we will follow the procedure described in
Table~\ref{tab:seqalgo}.  We generate an initial design of $n_0=10$
points (five times the dimension of the factor space) using a maximin
LHS (Latin Hypercube Sampling)\footnote{ More precisely, we use
  Matlab's \texttt{lhsdesign()} function to select the best design
  according to the maximin criterion among $10^4$ randomly generated
  LHS designs.} on $[-6;6] \times [-6;6]$.  We choose a Monte Carlo sample of size
$m=30000$. Since the true probability of failure is approximately
$\alpha = 0.4\%$ in this example, the coefficient of variation
for~$\alpha_m$ is $1 / \sqrt{m\, \alpha} \approx 9\%$. The same
initial design and Monte Carlo sample are used for all methods.

A Gaussian process with constant unknown mean and a Mat\'ern covariance
function is used as our prior information about $f$. The parameters of
the Mat\'ern covariance functions are estimated on the initial design by
REML \citep[see, e.g.][]{Ste99}. 
% We obtain $\sigma^2\approx 30$,
% $\nu\approx 3$ and $\rho_1 \approx \rho_2 \approx 12$, which corresponds
% to a fairly smooth prior. 
In this experiment, we follow the common practice of re-estimating the
parameters of the covariance function during the sequential strategy,
but only once every ten iterations to save some computation time.
% choose to re-estimate
% the parameters of the covariance once every ten iterations (after each new evaluation.

The probability of failure is estimated by~(\ref{eq:estimator1}). To
evaluate the rate of convergence, we compute the number $n_{\gamma}$ of
iterations that must be performed using a given strategy to observe a
stabilization of the relative error of estimation within an interval of
length~$2\gamma$:
$$
n_{\gamma} = \min \left\{n\geq 0; \forall k\geq n, \frac{\abs{\ahat_{n_0+k}-\alpha_m}}{\alpha_m}
  < \gamma \right\}\,.
$$
All the available sequential strategies are run 100 times, with
different initial designs and Monte Carlo samples. The results for
$\gamma = 0.10$, $\gamma = 0.03$ and $\gamma = 0.01$ are summarized in
Table~\ref{tab:result-four-branch}. We shall consider that~$n_{0.1}$
provides a measure of the performance of the strategy in the ``initial
phase'', where a rough estimate of~$\alpha$ is to be found,
whereas~$n_{0.03}$ and~$n_{0.01}$ measure the performance in the
``refinement phase''.

The four variants of the SUR strategy (see
Table~\ref{tab:expr-sur-crit}) have been run with~$Q = 12$ and
either~$m_0 = 10$ or~$m_0 = 500$. The performance are similar for all
four variants and for both values of~$m_0$. It appears, however, that
the criterions~$\JSUR{1}$ and~$\JSUR2{2}$ (i.e., the criterions given
directly by Proposition~\ref{prop:ub}) are slightly better
than~$\JSUR{3}$ and~$\JSUR{4}$; this will be confirmed by the
simulations of Section~\ref{sec:num:average}. It also seems that the
SUR algorithm is slightly slower to obtain a rough estimate of the
probability of failure when~$m_0$ is very small, but performs very
well in the refinement phase. (Note that~$m_0 = 10$ is a drastic
pruning for a sample of size~$m = 30000$.)

\begin{figure}[ht!]
   \includegraphics[width = 4cm]{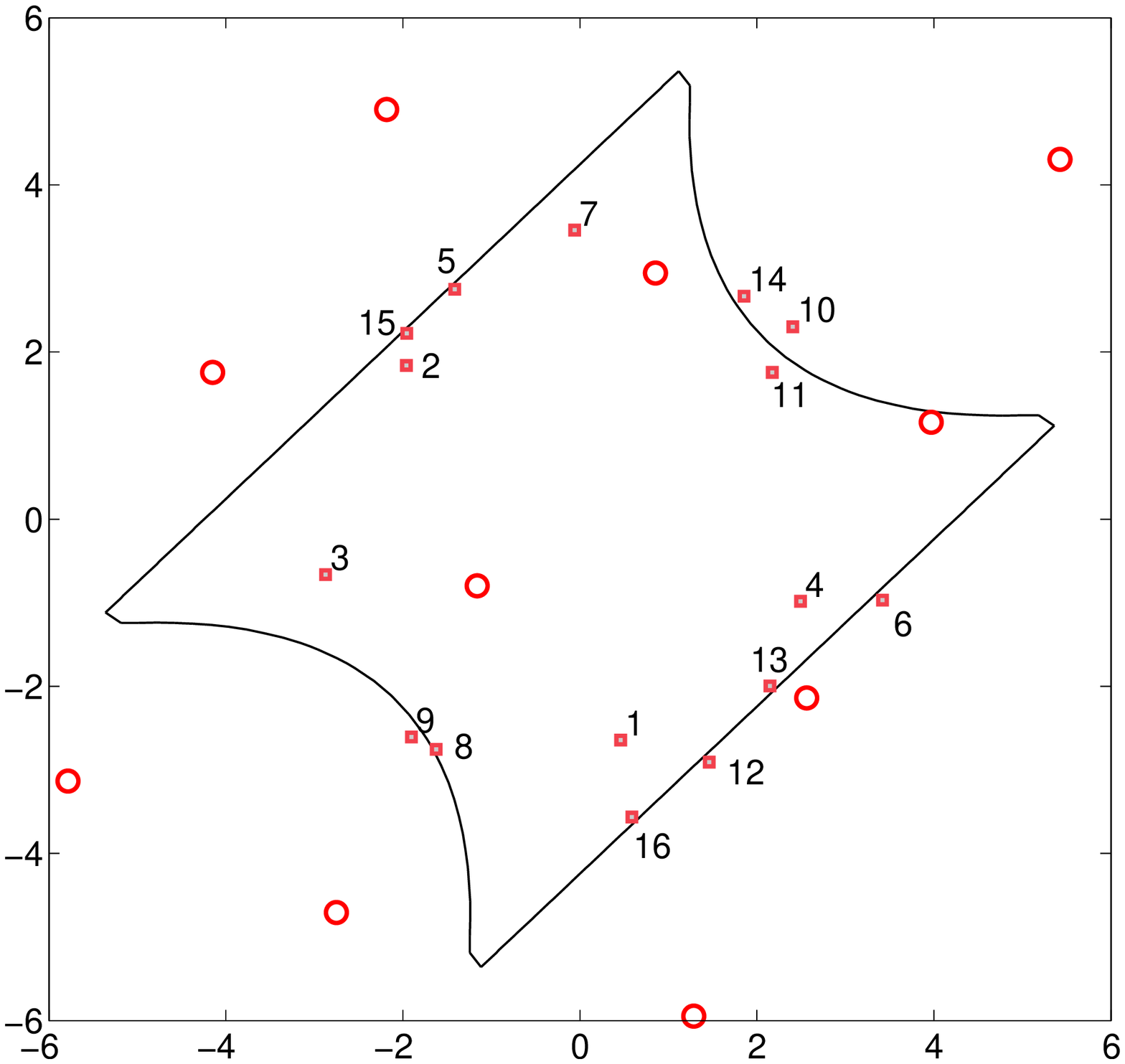}
   \includegraphics[width = 4cm]{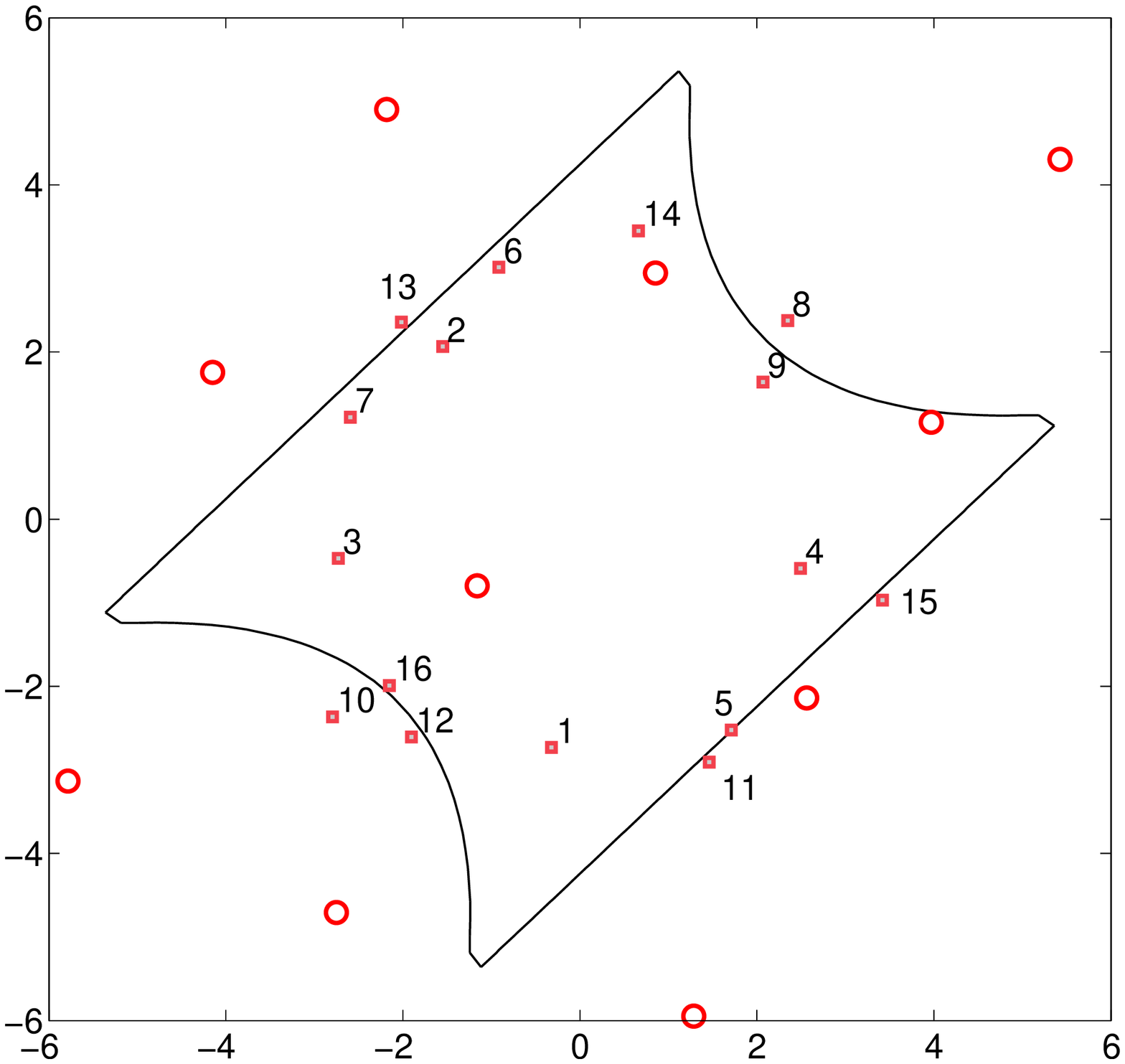}
   \includegraphics[width = 4cm]{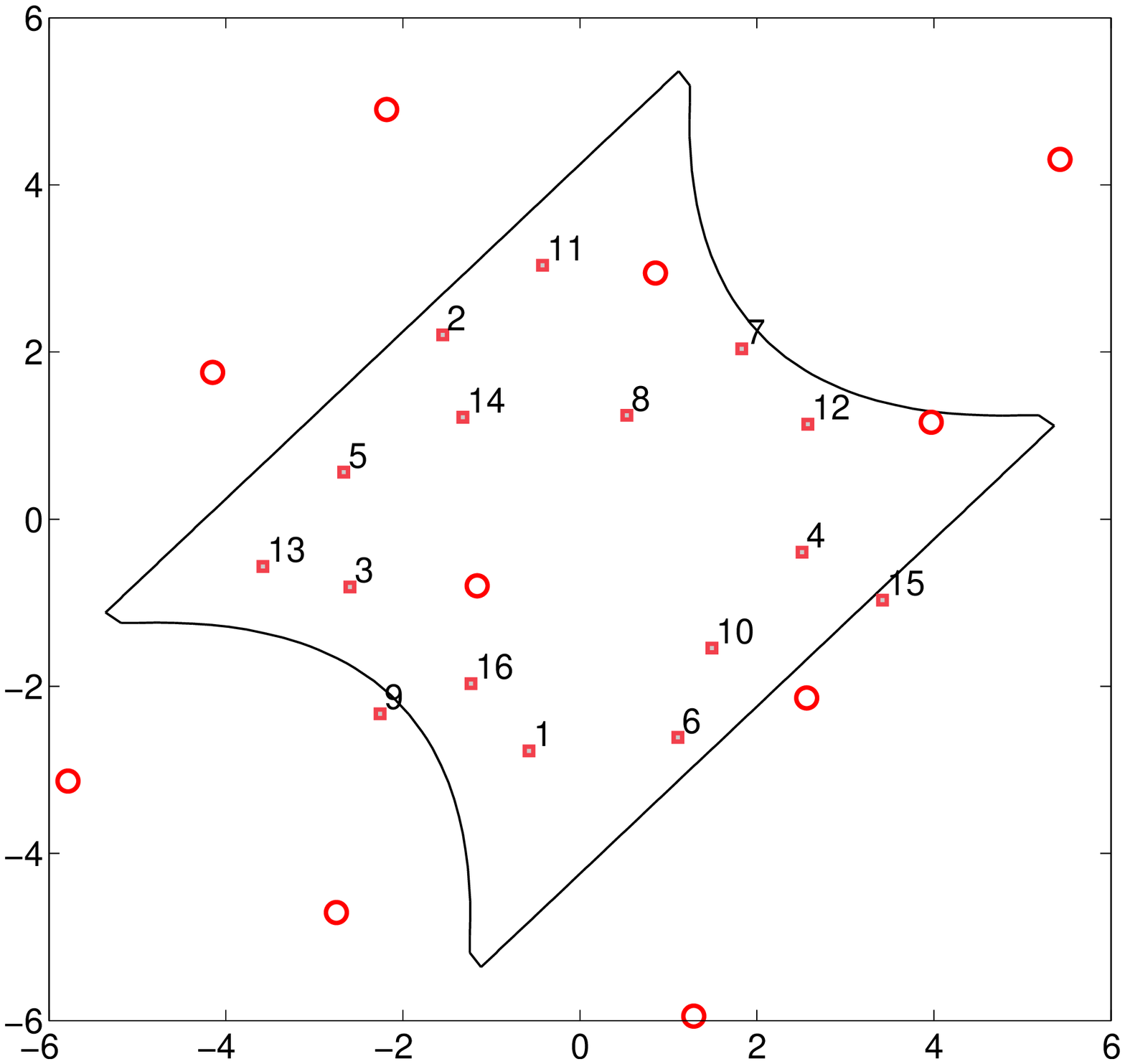}
   \caption{The first $16$ points (squares) evaluated using sampling
     criterion $\JSUR{1}$ (left), $\JTIMSE{}$ with
     $\sigma_{\varepsilon}^2=0.1$ (middle), $\JTIMSE{}$ with
     $\sigma_{\varepsilon}^2=1$ (right). Numbers near squares indicate the order of
     evaluation. The location of the $n_0=10$ points of the initial design are
     indicated by circles. }
   \label{fig:5.2res}
\end{figure}

The tIMSE strategy has been run for three different values of its
tuning parameter~$\sigma_{\varepsilon}^2$, using the pruning scheme
with~$m_0 = 500$. The best performance is obtained
for~$\sigma_\varepsilon^2 \approx 0$, and is almost as good as the
performance of SUR stragies with the same value of~$m0$ (a small loss
of performance, of about one evaluation on average, can be noticed in
the refinement phase).  Note that the required accuracy was not
reached after~200 iterations in 17\% of the runs for
$\sigma_\varepsilon^2 = 1$. In fact, the tIMSE strategy tends to
behave like a space-filling strategy in this
case. Figure~\ref{fig:5.2res} shows the points that have been
evaluated in three cases: the evaluations are less concentrated on the
boundary between the safe and the failure region when
$\sigma_\varepsilon^2 = 1$.

Finally, the results obtained for~$\JRB{}$ and~$\JECH{}$ indicate that
the corresponding strategies are clearly less efficient in the
``initial phase'' than strategies based on~$\JSUR{1}$
or~$\JSUR{2}$. For~$\gamma = 0.1$, the average loss with respect
to~$\JSUR{1}$ is between approximately $0.9$~evaluations for the best
case (criterion $\JRB{}$ with $\delta=2$, $\kappa=2$) and~$3.9$
evaluations for the worst case. For~$\gamma = 0.03$, the loss is
between~$1.4$ evaluations (also for (criterion $\JRB{}$ with
$\delta=2$, $\kappa=2$) and $3.5$~evaluations. This loss of efficiency
can also be observed very clearly on the $90^{\text{th}}$ percentile
in the inital phase. Criterion~$\JRB{}$ seems to perform best with
$\delta=2$ and $\kappa=2$ in this experiment, but this will not be
confirmed by the simulations of Section~\ref{sec:num:average}. Tuning
the parameters of this criterion for the estimation of a probability
of failure does not seem to be an easy task.

\begin{table}[htb]
  \centering
  \caption{Comparison of the convergence to $\alpha_{m}$ in the
    benchmark example Section~\ref{sec:num:structural} for different
    sampling strategies. The first number (bold text) is the average
    value of~$n_\gamma$ over 100 runs. The numbers between brackets
    indicate the $10^{\text{th}}$ and~$90^{\text{th}}$ percentile.}
  \label{tab:result-four-branch}
  \begin{tabular}{|l|l|l|l|l|}
    \hline 
    criterion   & parameters & $\gamma = 0.10$ & $\gamma = 0.03$ & $\gamma = 0.01$\\
    \hline
    $\JSUR{1}$  & $m_0= 500$  & \textbf{16.1} [10--22] & \textbf{25.7} [17--35] & \textbf{36.0} [26--48] \\ 
                & $m_0= 10$   & \textbf{19.4} [11--28] & \textbf{28.1} [19--38] & \textbf{35.4} [26--44] \\
    \hline   
    $\JSUR{2}$  & $m_0= 500$  & \textbf{16.4} [10--24] & \textbf{25.7} [19--33] & \textbf{35.5} [25--45] \\ 
                & $m_0= 10$   & \textbf{20.0} [11--30] & \textbf{28.3} [20--39] & \textbf{35.3} [26--44] \\
    \hline
    $\JSUR{3}$  & $m_0= 500$  & \textbf{18.2} [10--27] & \textbf{26.9} [18--37] & \textbf{35.9} [27--46] \\ 
                & $m_0= 10$   & \textbf{20.1} [11--30] & \textbf{28.0} [20--36] & \textbf{35.2} [25--44] \\
    \hline
    $\JSUR{4}$  & $m_0= 500$  & \textbf{17.2} [10--28] & \textbf{26.5} [20--36] & \textbf{35.2} [25--45] \\ 
                & $m_0= 10$   & \textbf{21.4} [13--30] & \textbf{28.9} [20--38] & \textbf{35.5} [27--44] \\
    \hline
    $\JTIMSE{}$ & $\sigma_{\varepsilon}^2=10^{-6}$ & \textbf{16.6} [10--23] & \textbf{26.5} [19--36] & \textbf{37.3} [28--49] \\
                & $\sigma_{\varepsilon}^2=0.1$ & \textbf{15.9} [10--22] & \textbf{29.1} [19--43] & \textbf{50.5} [30--79] \\
                & $\sigma_{\varepsilon}^2=1$ & \textbf{21.7} [11--31] & \textbf{52.4} [31--85] & \textbf{79.5} [42--133]${}^{ (*)}$\\
    \hline
    $\JECH{}$   & -- & \textbf{21.0} [11--31] & \textbf{29.2} [21--39] & \textbf{36.4} [28--44] \\ 
    \hline
    $\JRB{}$    & $\delta=1$, $\kappa=0.5$  & \textbf{18.7} [10--27] & \textbf{27.5} [20--35] & \textbf{36.6} [27--44] \\
                & $\delta=1$, $\kappa=2.0$  & \textbf{18.9} [11--28] & \textbf{28.3} [21--35] & \textbf{37.7} [30--45] \\
                & $\delta=2$, $\kappa=0.5$  & \textbf{17.6} [10--24] & \textbf{27.6} [20--34] & \textbf{37.1} [29--45] \\
                & $\delta=2$, $\kappa=2.0$  & \textbf{17.0} [10--21] & \textbf{27.1} [20--34] & \textbf{36.8} [29--44] \\
    \hline
  \end{tabular}
  \smallbreak
  {\small (*) The required accuracy was not reached after~200 iterations in 17\% of the runs}
\end{table}

\subsection{Average performance on sample paths of a Gaussian process}
\label{sec:num:average}

This section provides a comparison of all the criteria introduced or
recalled in this paper, on the basis of their average performance on the
sample paths of a zero-mean Gaussian process defined on~$\XX = \left[ 0,
  1 \right]^d$, for $d \in \left\{ 1, 2, 3 \right\}$. In all
experiments, the same covariance function is used for the generation of
the sample paths and for the computation of the sampling criteria.  We
have considered isotropic Mat\'ern covariance functions, whose
parameters are given in Table~\ref{tab:ap:cov}. An initial maximin LHS
design of size~$n_0$ (also given in the table) is used: note that the
value of~$n$ reported on the $x$-axis of Figures~\ref{fig:ap:sur}--\ref{fig:ap:all} is the total number
of evaluations, including the initial design.

\begin{table}[tbp]
  \caption{Size of the initial design and covariance parameters for 
    the experiments of Section~\ref{sec:num:average}. The 
    parametrization of the Mat\'ern covariance function used here is defined 
    in Appendix~\ref{sec:matern}.} 
  \label{tab:ap:cov}
  \centering
  \begin{tabular}{|c|c|c|c|c|}
    \hline
    $d$ & $n_0$ & $\sigma^2$ & $\nu$ & $\rho$  
    \\ \hline
    $1$ & $3$   & $1.0$     & $2.0$  & $0.100$ \\
    $2$ & $10$  & $1.0$     & $2.0$  & $0.252$ \\
    $3$ & $15$  & $1.0$     & $2.0$  & $0.363$ 
    \\ \hline
  \end{tabular}
\end{table}

The $d$ input variables are assumed to be independent and uniformly
distributed on~$[0,1]$, i.e., $\PX$ is the uniform distribution
on~$\XX$. An $m$-sample $Y_1$, \ldots, $Y_m$ from~$\PX$ is drawn one
and for all, and used both for the approximation of integrals (in SUR
and tIMSE criteria) and for the discrete search of the next sampling
point (for all criteria). We take $m = 500$ and use the same MC
sample for all criteria in a given dimension~$d$. 

\def\rMSE{\mathrm{rMSE}}

We adopt the meta-estimation framework as described in
Section~\ref{sec:SUR-discrete}; in other words, our goal is to
estimate the MC estimator~$\alpha_m$. We choose to adjust the
threshold~$u$ in order to have $\alpha_m = 0.02$ for all sample paths
(note that, as a consequence, there are exactly $m \alpha_m = 10$
points in the failure region) and we measure the performance of a
strategy after $n$ evaluations by its relative mean-square error (MSE)
expressed in decibels (dB):
\begin{equation*}
  \rMSE \;:=\; 
  10\, \log_{10} \left(\;
    \frac{1}{L}\,
    \sum_{l=1}^L
    \frac{ \left( \hat{\alpha}^{(l)}_{m,n} - \alpha_m \right)^2
    }{\alpha^2_m} \;\right)\,,
\end{equation*}
where $\hat{\alpha}^{(l)}_{m,n} = \frac{1}{m} \sum_{j=1}^m
p^{(l)}_n(Y_j)$ is the posterior mean of the MC estimator~$\alpha_m$
after $n$ evaluations on the $l^{\text{th}}$ simulated sample path ($L
= 4000$).

We use a sequential maximin strategy as a reference in all of our
experiments. This simple space-filling strategy is defined by $X_{n+1}
= \argmax_j\, \min_{1 \le i \le n} \left| Y_j - X_i \right|$, where
the argmax runs over all indices~$j$ such that~$Y_j \not\in \left\{
  X_1, \ldots, X_n \right\}$. Note that this strategy does not depend
on the choice of a Gaussian process model.

Our first experiment (Figure~\ref{fig:ap:sur}~) provides a comparison
of the four SUR strategies proposed in
Section~\ref{sec:SUR:upper-bounds}. It appears that all of them
perform roughly the same when compared to the reference strategy. A
closer look, however, reveals that the strategies~$\JSUR{1}$
and~$\JSUR{2}$ provided by Proposition~\ref{prop:ub} perform slightly
better than the other two (noticeably so in the case $d=3$).

The performance of the tIMSE strategy is shown on
Figure~\ref{fig:ap:tIMSE} for several value of its tuning parameter
$\se^2$ (other values, not shown here, have been tried as well). It is
clear that the performance of this strategy improves when $\se^2$ goes
to zero, whatever the dimension.

The performance of the strategy based on $\JRB{\kappa,\delta}$ is
shown on Figure~\ref{fig:ap:rb} for several values of its
parameters. It appears that the criterion proposed by
\citet{bichon:2008}, which corresponds to $\delta = 1$, performs
better than the one proposed by \citet{ranjan:2008}, which corresponds
to $\delta = 2$, for the same value of~$\kappa$. Moreover, the
value~$\kappa = 0.5$ has been found in our experiments to produce the
best results.

Figure~\ref{fig:ap:m0} illustrates that the loss of performance
associated to the ``pruning trick'' introduced in
Section~\ref{sec:SUR:impl} can be negligible if the size~$m_0$ of the
pruned MC sample is large enough (here, $m_0$ has been taken equal
to~$50$). In practice, the value of~$m_0$ should be chosen small
enough to keep the overhead of the sequential strategy reasonable---in
other words, large values of~$m_0$ should only be used for very
complex computer codes.

Finally, a comparison involving the best strategy obtained in each
category is presented on Figure~\ref{fig:ap:all}. The best result is
consistently obtained with the SUR strategy based on~$\JSUR{1,n}$. The
tIMSE strategy with $\se^2 \approx 0$ provides results which are
almost as good. Note that both strategies are one-step lookahead
strategies based on the approximation of the risk by an integral
criterion, which makes them rather expensive to compute. Simpler
strategies based on the marginal distribution (criteria $\JRB{n}$ and
$\JECH{n}$) provide interesting alternatives for moderately expensive
computer codes: their performances, although not as good as those of
one-step lookahead criterions, are still much better than that of the
reference space-filling strategy.

\def\titleCloserToFig{\vspace{-1ex}}
\def\reduceSpacingAfterFig{\vspace{0ex}}
\def\legend#1{{\scriptsize\emph{(#1)}}}

\begin{figure}[p]

\centering

%%%%%%%%%%%%%%%%%%%%%%%%   
%%%%%% ap-sur.tex %%%%%%
%%%%%%%%%%%%%%%%%%%%%%%%   

\begin{psfrags}%
\psfragscanon%
%
% text strings:
\psfrag{s10}[b][b]{\color[rgb]{0,0,0}\setlength{\tabcolsep}{0pt}\begin{tabular}{c}rMSE (dB)\end{tabular}}%
\psfrag{s11}[t][t]{\color[rgb]{0,0,0}\setlength{\tabcolsep}{0pt}\begin{tabular}{c}n\end{tabular}}%
\psfrag{s12}[b][b]{\color[rgb]{0,0,0}\setlength{\tabcolsep}{0pt}\begin{tabular}{c}rMSE (dB)\end{tabular}}%
\psfrag{s13}[t][t]{\color[rgb]{0,0,0}\setlength{\tabcolsep}{0pt}\begin{tabular}{c}n\end{tabular}}%
\psfrag{s14}[b][b]{\color[rgb]{0,0,0}\setlength{\tabcolsep}{0pt}\begin{tabular}{c}rMSE (dB)\end{tabular}}%
\psfrag{s15}[t][t]{\color[rgb]{0,0,0}\setlength{\tabcolsep}{0pt}\begin{tabular}{c}n\end{tabular}}%
\psfrag{s17}[lt][lt]{\hspace{-2.5ex}\begin{minipage}{5cm}%
    \scriptsize 
    \begin{tabular}{ll}
      $\JSUR{1}$ & black solid line\\
      $\JSUR{2}$ & gray solid line\\
      $\JSUR3$   & gray mixed line\\
      $\JSUR4$   & black mixed line\\
      ref.       & black dashed line
    \end{tabular}
    \vspace{2ex}\\
    \begin{tabular}{ll}
      Upper-left:  & $d=1$ \\
      Upper-right: & $d=2$ \\
      Lower-left:  & $d=3$ \\     
    \end{tabular}
  \end{minipage}}
%
% xticklabels:
%
\psfrag{x01}[t][t]{0}%
\psfrag{x02}[t][t]{0.5}%
\psfrag{x03}[t][t]{1}%
\psfrag{x04}[t][t]{20}%
\psfrag{x05}[t][t]{40}%
\psfrag{x06}[t][t]{60}%
\psfrag{x07}[t][t]{80}%
\psfrag{x08}[t][t]{100}%
\psfrag{x09}[t][t]{20}%
\psfrag{x10}[t][t]{40}%
\psfrag{x11}[t][t]{60}%
\psfrag{x12}[t][t]{80}%
\psfrag{x13}[t][t]{10}%
\psfrag{x14}[t][t]{20}%
\psfrag{x15}[t][t]{30}%
%
% yticklabels:
%
\psfrag{v01}[r][r]{0}%
\psfrag{v02}[r][r]{0.5}%
\psfrag{v03}[r][r]{1}%
\psfrag{v04}[r][r]{-25}%
\psfrag{v05}[r][r]{-20}%
\psfrag{v06}[r][r]{-15}%
\psfrag{v07}[r][r]{-10}%
\psfrag{v08}[r][r]{-5}%
\psfrag{v09}[r][r]{-30}%
\psfrag{v10}[r][r]{-20}%
\psfrag{v11}[r][r]{-10}%
\psfrag{v12}[r][r]{0}%
\psfrag{v13}[r][r]{-30}%
\psfrag{v14}[r][r]{-20}%
\psfrag{v15}[r][r]{-10}%
\psfrag{v16}[r][r]{0}%
\psfrag{v17}[r][r]{10}%
% Figure:
\resizebox{12.2cm}{!}{\includegraphics{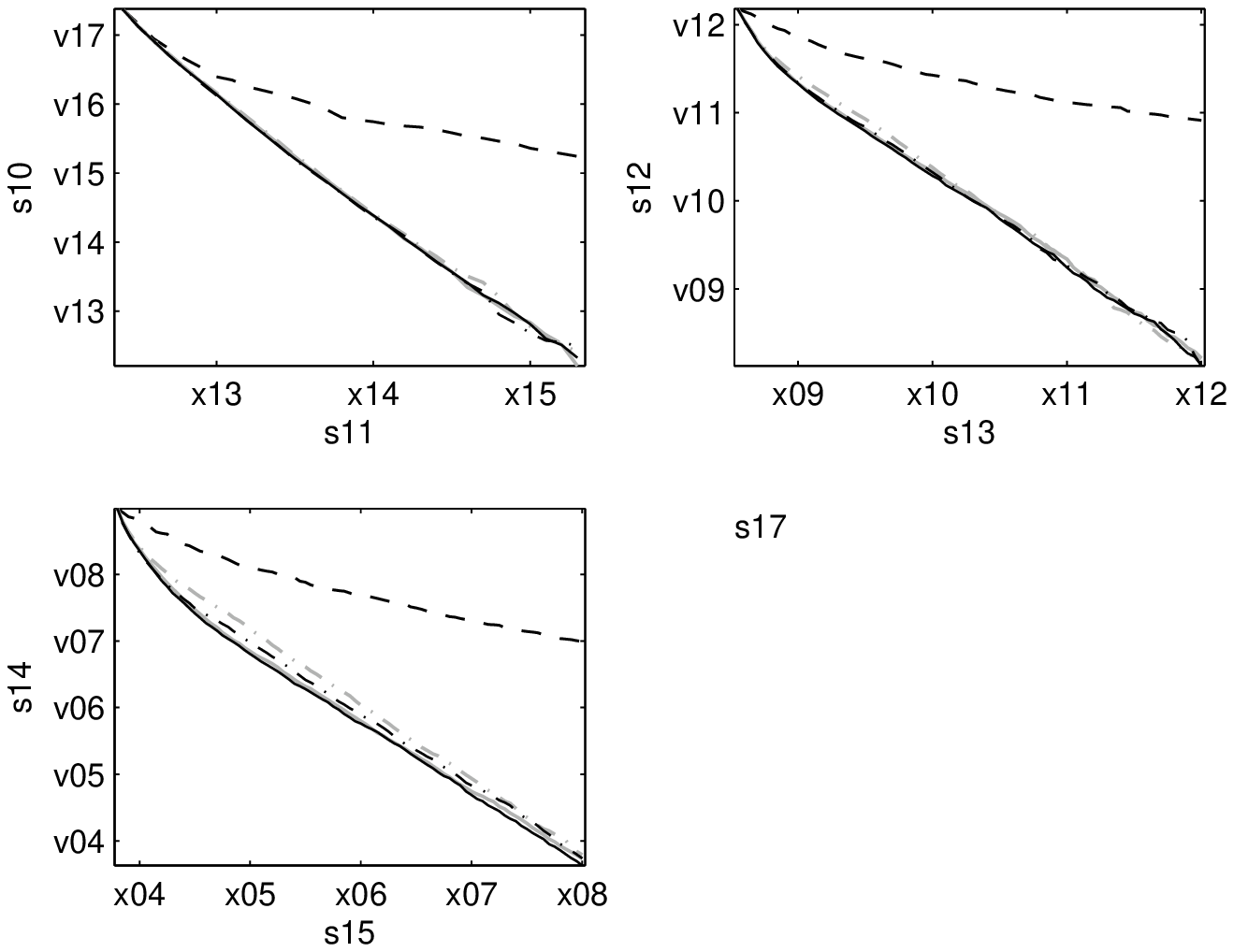}}%
\end{psfrags}

  \titleCloserToFig
  \caption{Relative MSE performance of several SUR strategies.}
  \reduceSpacingAfterFig
  \label{fig:ap:sur}  

  \vspace{.5cm}

%%%%%%%%%%%%%%%%%%%%%%%%%%%%%%%
%%%%%%%% ap-tIMSE.tex %%%%%%%%%
%%%%%%%%%%%%%%%%%%%%%%%%%%%%%%%

\begin{psfrags}%
\psfragscanon%
%
% text strings:
\psfrag{s10}[b][b]{\color[rgb]{0,0,0}\setlength{\tabcolsep}{0pt}\begin{tabular}{c}rMSE (dB)\end{tabular}}%
\psfrag{s11}[t][t]{\color[rgb]{0,0,0}\setlength{\tabcolsep}{0pt}\begin{tabular}{c}n\end{tabular}}%
\psfrag{s12}[b][b]{\color[rgb]{0,0,0}\setlength{\tabcolsep}{0pt}\begin{tabular}{c}rMSE (dB)\end{tabular}}%
\psfrag{s13}[t][t]{\color[rgb]{0,0,0}\setlength{\tabcolsep}{0pt}\begin{tabular}{c}n\end{tabular}}%
\psfrag{s14}[b][b]{\color[rgb]{0,0,0}\setlength{\tabcolsep}{0pt}\begin{tabular}{c}rMSE (dB)\end{tabular}}%
\psfrag{s15}[t][t]{\color[rgb]{0,0,0}\setlength{\tabcolsep}{0pt}\begin{tabular}{c}n\end{tabular}}%
\psfrag{s17}[lt][lt]{\hspace{-2.5ex}\begin{minipage}{5cm}%
    \scriptsize 
    \begin{tabular}{ll}
      $\se^2 = 10^{-6}$ & black solid line\\
      $\se^2 = 0.1$ & gray solid line\\
      $\se^2 = 1$ & black mixed line\\
      ref. & black dashed line
    \end{tabular}
    \vspace{2ex}\\
    \begin{tabular}{ll}
      Upper-left:  & $d=1$ \\
      Upper-right: & $d=2$ \\
      Lower-left:  & $d=3$ \\
    \end{tabular}
  \end{minipage}}

%
% xticklabels:
%
\psfrag{x01}[t][t]{0}%
\psfrag{x02}[t][t]{0.5}%
\psfrag{x03}[t][t]{1}%
\psfrag{x04}[t][t]{20}%
\psfrag{x05}[t][t]{40}%
\psfrag{x06}[t][t]{60}%
\psfrag{x07}[t][t]{80}%
\psfrag{x08}[t][t]{100}%
\psfrag{x09}[t][t]{20}%
\psfrag{x10}[t][t]{40}%
\psfrag{x11}[t][t]{60}%
\psfrag{x12}[t][t]{80}%
\psfrag{x13}[t][t]{10}%
\psfrag{x14}[t][t]{20}%
\psfrag{x15}[t][t]{30}%
%
% yticklabels:
\psfrag{v01}[r][r]{0}%
\psfrag{v02}[r][r]{0.5}%
\psfrag{v03}[r][r]{1}%
\psfrag{v04}[r][r]{-25}%
\psfrag{v05}[r][r]{-20}%
\psfrag{v06}[r][r]{-15}%
\psfrag{v07}[r][r]{-10}%
\psfrag{v08}[r][r]{-5}%
\psfrag{v09}[r][r]{-30}%
\psfrag{v10}[r][r]{-20}%
\psfrag{v11}[r][r]{-10}%
\psfrag{v12}[r][r]{0}%
\psfrag{v13}[r][r]{-30}%
\psfrag{v14}[r][r]{-20}%
\psfrag{v15}[r][r]{-10}%
\psfrag{v16}[r][r]{0}%
\psfrag{v17}[r][r]{10}%
%
% Figure:
\resizebox{12.2cm}{!}{\includegraphics{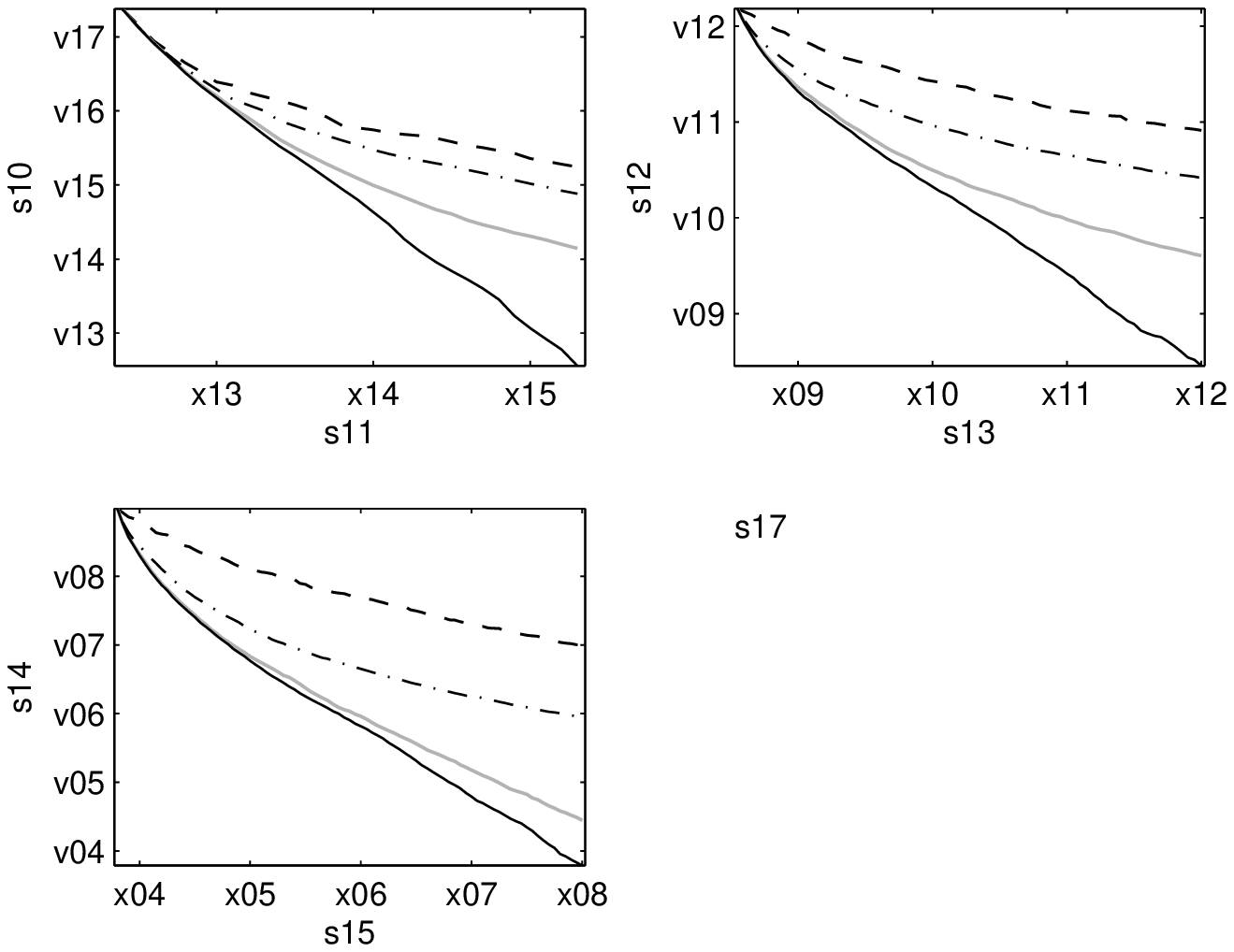}}%
\end{psfrags}%

  \titleCloserToFig
  \caption{Relative MSE performance of the tIMSE strategy for several
    values of its parameter.}
  \reduceSpacingAfterFig
  \label{fig:ap:tIMSE}

\end{figure}

\begin{figure}[p]
  
\centering

%%%%%%%%%%%%%%%%%%%%%%%%%%%%
%%%%%%%%% ap-rb.tex %%%%%%%%
%%%%%%%%%%%%%%%%%%%%%%%%%%%%

\begin{psfrags}%
\psfragscanon%
%
% text strings:
\psfrag{s10}[b][b]{\color[rgb]{0,0,0}\setlength{\tabcolsep}{0pt}\begin{tabular}{c}rMSE (dB)\end{tabular}}%
\psfrag{s11}[t][t]{\color[rgb]{0,0,0}\setlength{\tabcolsep}{0pt}\begin{tabular}{c}n\end{tabular}}%
\psfrag{s12}[b][b]{\color[rgb]{0,0,0}\setlength{\tabcolsep}{0pt}\begin{tabular}{c}rMSE (dB)\end{tabular}}%
\psfrag{s13}[t][t]{\color[rgb]{0,0,0}\setlength{\tabcolsep}{0pt}\begin{tabular}{c}n\end{tabular}}%
\psfrag{s14}[b][b]{\color[rgb]{0,0,0}\setlength{\tabcolsep}{0pt}\begin{tabular}{c}rMSE (dB)\end{tabular}}%
\psfrag{s15}[t][t]{\color[rgb]{0,0,0}\setlength{\tabcolsep}{0pt}\begin{tabular}{c}n\end{tabular}}%
\psfrag{s17}[lt][lt]{\hspace{-2.5ex}\begin{minipage}{5cm}%
    \scriptsize 
    \begin{tabular}{lll}
      $\kappa=0.5$ & $\delta=1$ & black solid line\\
      $\kappa=0.5$ & $\delta=2$ & gray solid line\\
      $\kappa=2$   & $\delta=1$ & black mixed line\\
      $\kappa=2$   & $\delta=2$ & gray mixed line\\
      ref. & & black dashed line
    \end{tabular}
    \vspace{2ex}\\
    \begin{tabular}{ll}
      Upper-left:  & $d=1$ \\
      Upper-right: & $d=2$ \\
      Lower-left:  & $d=3$   
    \end{tabular}
  \end{minipage}}

%
% xticklabels:
%
\psfrag{x01}[t][t]{0}%
\psfrag{x02}[t][t]{0.5}%
\psfrag{x03}[t][t]{1}%
\psfrag{x04}[t][t]{20}%
\psfrag{x05}[t][t]{40}%
\psfrag{x06}[t][t]{60}%
\psfrag{x07}[t][t]{80}%
\psfrag{x08}[t][t]{100}%
\psfrag{x09}[t][t]{20}%
\psfrag{x10}[t][t]{40}%
\psfrag{x11}[t][t]{60}%
\psfrag{x12}[t][t]{80}%
\psfrag{x13}[t][t]{10}%
\psfrag{x14}[t][t]{20}%
\psfrag{x15}[t][t]{30}%
%
% yticklabels:
%
\psfrag{v01}[r][r]{0}%
\psfrag{v02}[r][r]{0.5}%
\psfrag{v03}[r][r]{1}%
\psfrag{v04}[r][r]{-25}%
\psfrag{v05}[r][r]{-20}%
\psfrag{v06}[r][r]{-15}%
\psfrag{v07}[r][r]{-10}%
\psfrag{v08}[r][r]{-5}%
\psfrag{v09}[r][r]{-30}%
\psfrag{v10}[r][r]{-20}%
\psfrag{v11}[r][r]{-10}%
\psfrag{v12}[r][r]{0}%
\psfrag{v13}[r][r]{-30}%
\psfrag{v14}[r][r]{-20}%
\psfrag{v15}[r][r]{-10}%
\psfrag{v16}[r][r]{0}%
\psfrag{v17}[r][r]{10}%
%
% Figure:
\resizebox{12.2cm}{!}{\includegraphics{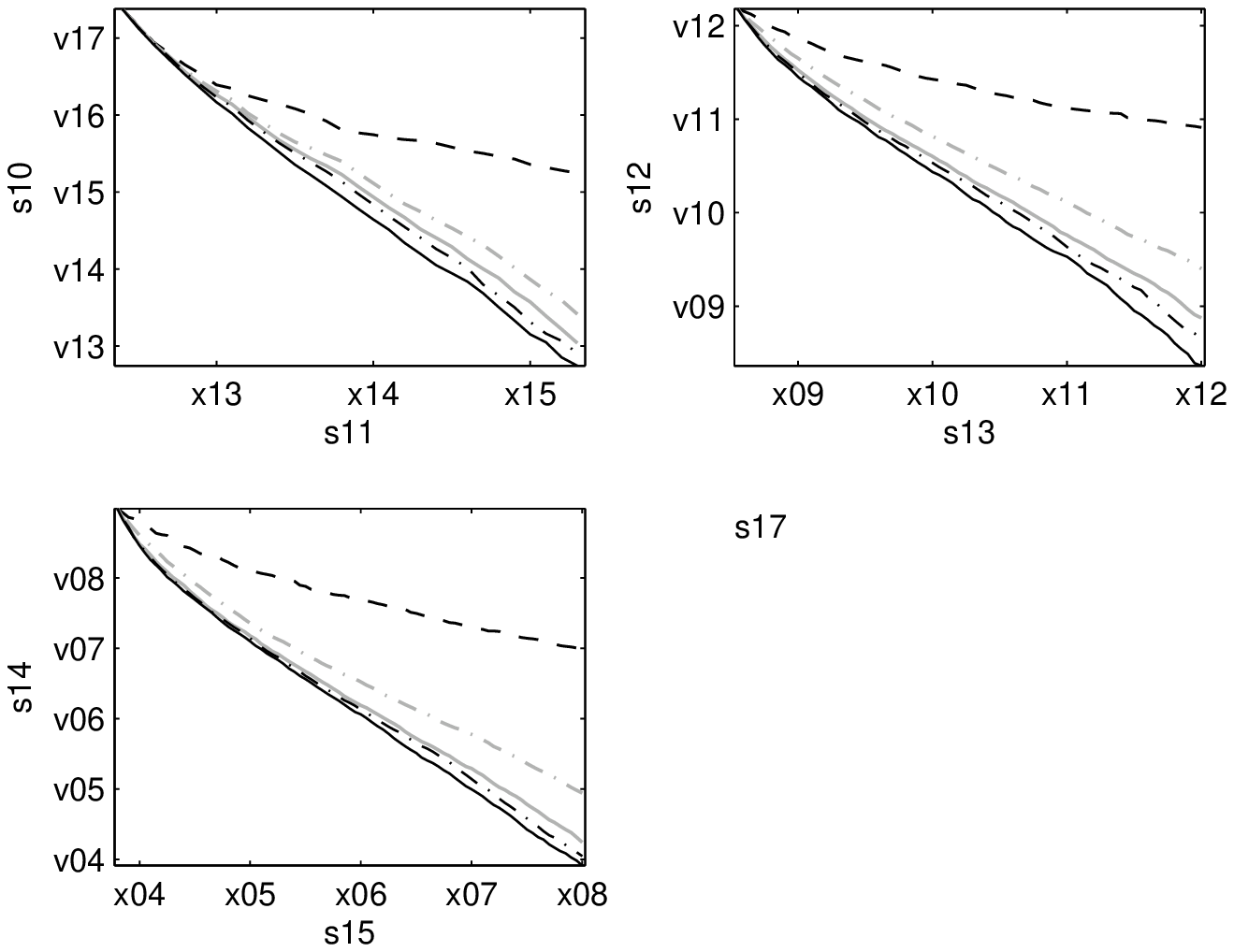}}%
\end{psfrags}%

  \titleCloserToFig
  \caption{Relative MSE performance of the $\JRB{}$ criterion, for
    several values of its parameters.}
  \reduceSpacingAfterFig
  \label{fig:ap:rb}

  \vspace{.5cm}

%%%%%%%%%%%%%%%%%%%%%%%%%%%%%%%
%%%%%%%%% ap-m0.tex %%%%%%%%%%%
%%%%%%%%%%%%%%%%%%%%%%%%%%%%%%%

\begin{psfrags}%
\psfragscanon%
%
% text strings:
\psfrag{s10}[b][b]{\color[rgb]{0,0,0}\setlength{\tabcolsep}{0pt}\begin{tabular}{c}rMSE (dB)\end{tabular}}%
\psfrag{s11}[t][t]{\color[rgb]{0,0,0}\setlength{\tabcolsep}{0pt}\begin{tabular}{c}n\end{tabular}}%
\psfrag{s12}[b][b]{\color[rgb]{0,0,0}\setlength{\tabcolsep}{0pt}\begin{tabular}{c}rMSE (dB)\end{tabular}}%
\psfrag{s13}[t][t]{\color[rgb]{0,0,0}\setlength{\tabcolsep}{0pt}\begin{tabular}{c}n\end{tabular}}%
\psfrag{s14}[b][b]{\color[rgb]{0,0,0}\setlength{\tabcolsep}{0pt}\begin{tabular}{c}rMSE (dB)\end{tabular}}%
\psfrag{s15}[t][t]{\color[rgb]{0,0,0}\setlength{\tabcolsep}{0pt}\begin{tabular}{c}n\end{tabular}}%
\psfrag{s17}[lt][lt]{\hspace{-2.5ex}\begin{minipage}{5cm}%
    \scriptsize 
    \begin{tabular}{ll}
      $\JSUR{1}$ & full line\\
      $\JSUR3$ & mixed line\\
      without pruning & black\\
      pruning $m_0 = 50$ & gray\\
      ref. & black dashed line
    \end{tabular}
    \vspace{2ex}\\
    \begin{tabular}{ll}
      Upper-left:  & $d=1$ \\
      Upper-right: & $d=2$ \\
      Lower-left:  & $d=3$ \\    
    \end{tabular}
  \end{minipage}}

%
% xticklabels:
%
\psfrag{x01}[t][t]{0}%
\psfrag{x02}[t][t]{0.5}%
\psfrag{x03}[t][t]{1}%
\psfrag{x04}[t][t]{20}%
\psfrag{x05}[t][t]{40}%
\psfrag{x06}[t][t]{60}%
\psfrag{x07}[t][t]{80}%
\psfrag{x08}[t][t]{100}%
\psfrag{x09}[t][t]{20}%
\psfrag{x10}[t][t]{40}%
\psfrag{x11}[t][t]{60}%
\psfrag{x12}[t][t]{80}%
\psfrag{x13}[t][t]{10}%
\psfrag{x14}[t][t]{20}%
\psfrag{x15}[t][t]{30}%
%
% yticklabels:
%
\psfrag{v01}[r][r]{0}%
\psfrag{v02}[r][r]{0.5}%
\psfrag{v03}[r][r]{1}%
\psfrag{v04}[r][r]{-25}%
\psfrag{v05}[r][r]{-20}%
\psfrag{v06}[r][r]{-15}%
\psfrag{v07}[r][r]{-10}%
\psfrag{v08}[r][r]{-5}%
\psfrag{v09}[r][r]{-30}%
\psfrag{v10}[r][r]{-20}%
\psfrag{v11}[r][r]{-10}%
\psfrag{v12}[r][r]{0}%
\psfrag{v13}[r][r]{-40}%
\psfrag{v14}[r][r]{-20}%
\psfrag{v15}[r][r]{0}%
%
% Figure:
\resizebox{12.2cm}{!}{\includegraphics{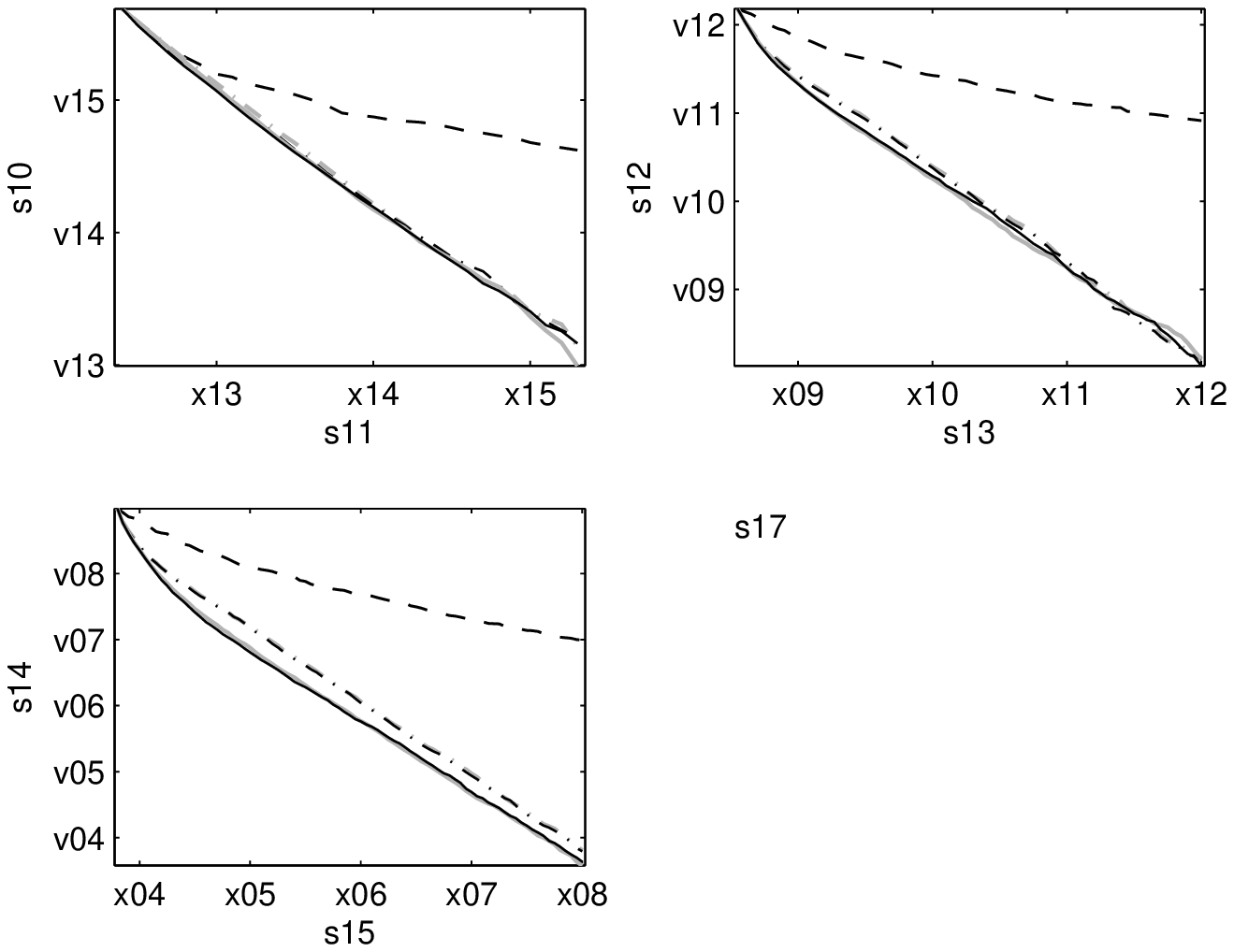}}%
\end{psfrags}%

  \titleCloserToFig
  \caption{Relative MSE performance of two SUR criteria, with and
    without the ``pruning trick'' described in 
    Section~\ref{sec:SUR:impl}. The black and gray lines are almost
    surimposed for each of the criterions $\JSUR{1}$ and~$\JSUR{3}$.}
  \reduceSpacingAfterFig
  \label{fig:ap:m0}

\end{figure}

\begin{figure}[t]

\centering

%%%%%%%%%%%%%%%%%%%%%%%%%%%%%%%%%%%
%%%%%%%%%% ap-all.tex %%%%%%%%%%%%%
%%%%%%%%%%%%%%%%%%%%%%%%%%%%%%%%%%%

\begin{psfrags}%
\psfragscanon%
%
% text strings:
\psfrag{s10}[b][b]{\color[rgb]{0,0,0}\setlength{\tabcolsep}{0pt}\begin{tabular}{c}rMSE (dB)\end{tabular}}%
\psfrag{s11}[t][t]{\color[rgb]{0,0,0}\setlength{\tabcolsep}{0pt}\begin{tabular}{c}n\end{tabular}}%
\psfrag{s12}[b][b]{\color[rgb]{0,0,0}\setlength{\tabcolsep}{0pt}\begin{tabular}{c}rMSE (dB)\end{tabular}}%
\psfrag{s13}[t][t]{\color[rgb]{0,0,0}\setlength{\tabcolsep}{0pt}\begin{tabular}{c}n\end{tabular}}%
\psfrag{s14}[b][b]{\color[rgb]{0,0,0}\setlength{\tabcolsep}{0pt}\begin{tabular}{c}rMSE (dB)\end{tabular}}%
\psfrag{s15}[t][t]{\color[rgb]{0,0,0}\setlength{\tabcolsep}{0pt}\begin{tabular}{c}n\end{tabular}}%
\psfrag{s17}[lt][lt]{\hspace{-2.5ex}\begin{minipage}{5cm}%
    \scriptsize 
    \begin{tabular}{ll}
      $\JSUR{1}$ & black solid line\\
      tIMSE ($\se^2 = 10^{-6}$) & gray solid line\\
      $\JRB{}$ ($\kappa=0.5$, $\delta =1$) & black mixed line\\
      $\JECH{}$ & gray mixed line\\
      ref. & black dashed line\\
    \end{tabular}
    \vspace{2ex}\\
    \begin{tabular}{ll}
      Upper-left:  & $d=1$ \\
      Upper-right: & $d=2$ \\
      Lower-left:  & $d=3$ \\    
    \end{tabular}
  \end{minipage}}

%
% xticklabels:
\psfrag{x01}[t][t]{0}%
\psfrag{x02}[t][t]{0.5}%
\psfrag{x03}[t][t]{1}%
\psfrag{x04}[t][t]{20}%
\psfrag{x05}[t][t]{40}%
\psfrag{x06}[t][t]{60}%
\psfrag{x07}[t][t]{80}%
\psfrag{x08}[t][t]{100}%
\psfrag{x09}[t][t]{20}%
\psfrag{x10}[t][t]{40}%
\psfrag{x11}[t][t]{60}%
\psfrag{x12}[t][t]{80}%
\psfrag{x13}[t][t]{10}%
\psfrag{x14}[t][t]{20}%
\psfrag{x15}[t][t]{30}%
%
% yticklabels:
\psfrag{v01}[r][r]{0}%
\psfrag{v02}[r][r]{0.5}%
\psfrag{v03}[r][r]{1}%
\psfrag{v04}[r][r]{-25}%
\psfrag{v05}[r][r]{-20}%
\psfrag{v06}[r][r]{-15}%
\psfrag{v07}[r][r]{-10}%
\psfrag{v08}[r][r]{-5}%
\psfrag{v09}[r][r]{-30}%
\psfrag{v10}[r][r]{-20}%
\psfrag{v11}[r][r]{-10}%
\psfrag{v12}[r][r]{0}%
\psfrag{v13}[r][r]{-30}%
\psfrag{v14}[r][r]{-20}%
\psfrag{v15}[r][r]{-10}%
\psfrag{v16}[r][r]{0}%
\psfrag{v17}[r][r]{10}%
%
% Figure:
\resizebox{12.2cm}{!}{\includegraphics{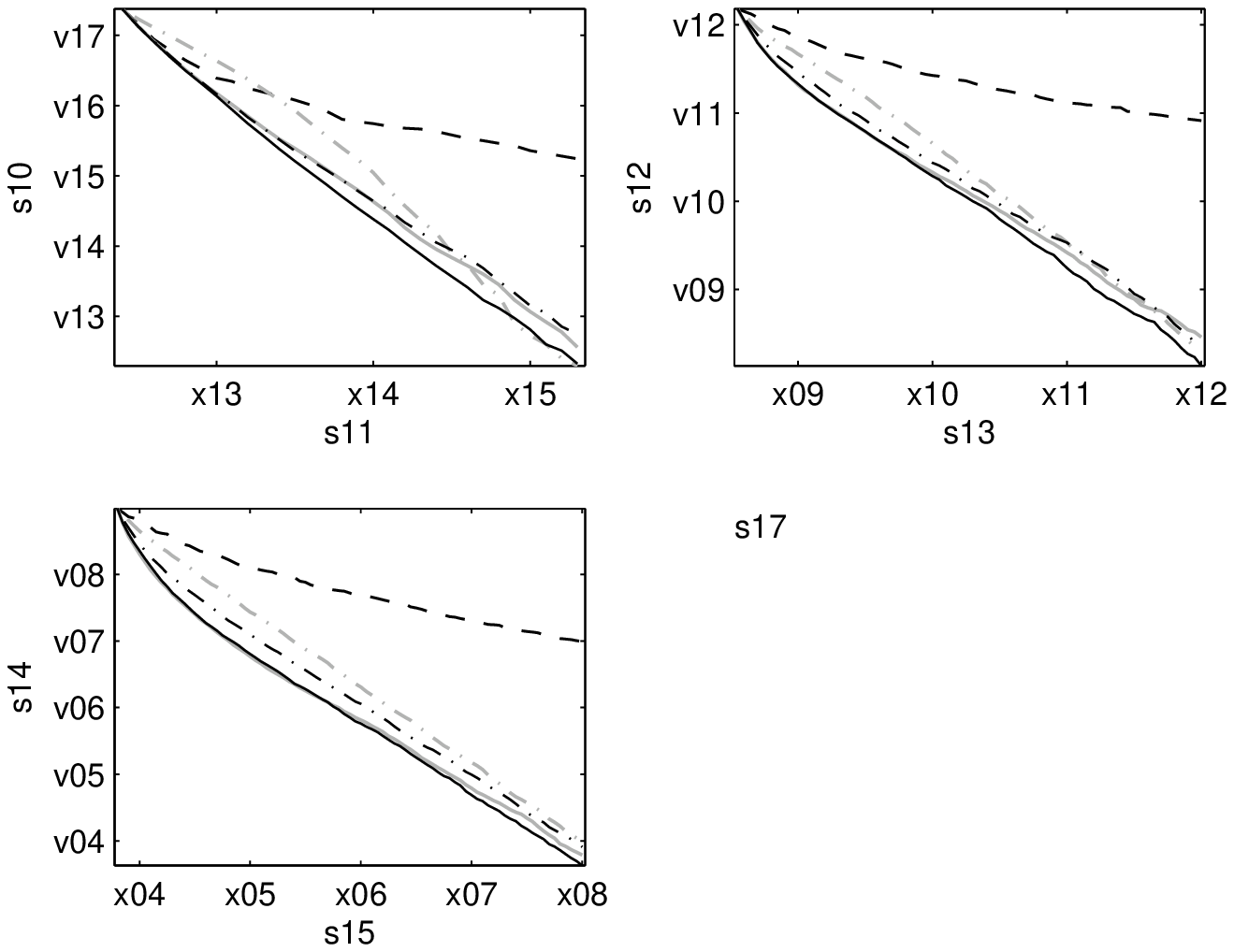}}%
\end{psfrags}%

  \titleCloserToFig
  \caption{Relative MSE performance the best strategy in each
    category.}
  \reduceSpacingAfterFig
  \label{fig:ap:all}
\end{figure}

\section{Concluding remarks}
\label{sec:discuss}

One of the main objectives of this paper was to present a synthetic
viewpoint on sequential strategies based on a Gaussian process model and
kriging for the estimation of a probability of failure. The starting
point of this presentation is a Bayesian decision-theoretic framework
from which the theoretical form of an optimal strategy for the
estimation of a probability of failure can be derived. Unfortunately,
the dynamic programming problem corresponding to this strategy is not
numerically tractable. It is nonetheless possible to derive from there the
ingredients of a sub-optimal strategy: the idea is to focus on
one-step lookahead suboptimal strategies, where the exact risk is
replaced by a substitute risk that accounts for the information gain
about~$\alpha$ expected from a new evaluation. We call such a strategy a
\emph{stepwise uncertainty reduction} (SUR) strategy. 
Our numerical experiments show that SUR strategies perform better, on average,
than the other strategies proposed in the literature. However, this comes
at a higher computational cost than strategies based only on marginal
distributions. The tIMSE sampling criterion, which seems to have a
convergence rate comparable to that of the SUR criterions when
$\se^2 \approx 0$, also has a high computational complexity.

In which situations can we say that the sequential strategies presented in this paper are
interesting alternatives to classical importance sampling methods for
estimating a probability of failure, for instance the subset sampling
method of \cite{au01:_estim}? In our opinion, beyond the obvious role
of the simulation budget~$N$, the answer to this
question depends on our capacity to elicit an appropriate prior. In the
example of Section~\ref{sec:num:structural}, as well as in many other
examples of the literature using Gaussian processes in the domain of
computer experiments, the prior is easy to choose because $\XX$ is a
low-dimensional space and $f$ tends to be smooth. Then, the plug-in
approach which consists in using ML or REML to estimate the parameters
of the covariance function of the Gaussian process after each new
evaluation is likely to succeed. If $\XX$ is high-dimensional and $f$ is
expensive to evaluate, difficulties arise. In particular, our sampling
strategies do not take into account our uncertain knowledge of the
covariance parameters, and there is no guarantee that ML estimation will do
well when the points are chosen by a sampling strategy that favors some localized
target region (the neighboorhood the frontier of the domain of failure
in this paper, but the question is equally relevant in the field 
optimization, for instance).
The difficult problem of deciding the size~$n_0$ of the initial design is crucial in this
connection. Fully Bayes procedures constitute a possible direction
for future research, as long as they don't introduce an unacceptable 
computational overhead. Whatever the route, we feel that
the robustness of Gaussian process-based sampling strategies with
respect to the procedure of estimation of the covariance parameters
should be addressed carefully in order to make these methods usable in
the industrial world.

\bigbreak

\footnotesize \noindent \textbf{Software.} We would like to draw the
reader's attention on the recently published package KrigInv
\citep{picheny:2011:kriginv} for the statistical computing environment
R \citep[see][]{hornik:2010:r}. This package provides an open source
(GPLv3) implementation of all the strategies proposed in this
paper. Please note that the simulation results presented in this paper
were not obtained using this package, that was not available at the
time of its writing.

\normalsize

\begin{acknowledgements}
  The research of Julien Bect, Ling Li and Emmanuel Vazquez was
  partially funded by the French \emph{Agence Nationale de la
    Recherche} (ANR) in the context of the project \textsc{opus}
  (ref. \textsc{anr-07-cis7-010}) and by the French \emph{p\^ole de
    comp\'etitivit\'e} \textsc{systematic} in the context of the
  project \textsc{csdl}. David Ginsbourger acknowledges support from
  the French \emph{Institut de Radioprotection et de Sûreté Nucléaire}
  (IRSN) and warmly thanks Yann Richet.
\end{acknowledgements}

\clearpage
\appendix
\noindent
{\Large Appendix}
\section{The Mat\'ern covariance}
\label{sec:matern}

The exponential covariance and the Mat\'ern covariance are among the
most conventionally used stationary covariances in the literature of
design and analysis of computer experiments.  The Mat\'ern covariance
class \citep{yaglom86a} offers the possibility to adjust the regularity
of~$\xi$ with a single parameter.  \cite{Ste99} advocates the use of the
following parametrization of the Mat\'ern function:
\begin{equation}
  \kappa_{\nu}(h) =  \frac{1}{2^{\nu-1}\Gamma(\nu)}\left(2\nu^{1/2}h\right)^\nu
  \mathcal{K}_\nu\left(2\nu^{1/2}h\right)\,,\quad h\in\RR
\end{equation}
where $\Gamma$ is the Gamma function and $\mathcal{K}_\nu$ is the modified
Bessel function of the second kind.  The parameter $\nu>0$ controls
regularity at the origin of the function. To model a real-valued
function $f$ defined over $\XX \subset \RR^d$, with $d\geq1$, we use the
following anisotropic form of the Mat\'ern covariance:
\begin{equation}
  \label{eq:materncov}
  k_{\theta}(x,y) = \sigma^2 \kappa_{\nu}\left( \sqrt{ \sum_{i=1}^d
      \frac{(x_{[i]} -y_{[i]})^2}{\rho_i^2}}\right)\,, \quad x,y\in\RR^d
\end{equation}
where $x_{[i]}, y_{[i]}$ denote the $i^{\rm th}$ coordinate of $x$ and
$y$, the positive scalar $\sigma^2$ is a variance parameter (we have
$k_{\theta}(x,x) = \sigma^2$), and the positive scalars $\rho_i$
represent scale or \emph{range} parameters of the covariance,
\emph{i.e.}, characteristic correlation lengths. Since $\sigma^2>0,
\nu>0, \rho_i>0$, $i =1,\ldots, d$, we can take the logarithm of these scalars,
and consider the vector of parameters $\theta=\{\log \sigma^2,
\log \nu, -\log \rho_1,\ldots, -\log\rho_d\}\in\RR^{d+2}$, which is a
practical parameterization when $\sigma^2,
\nu, \rho_i$, $i =1,\ldots, d$, need to be estimated from data.

\section{Proof of Proposition~\ref{prop:crit-RB}}
\label{sec:comput:rb}

a)\ Using the identity $\normCDF^{-1}(1-p) = -\, \normCDF^{-1}(p)$, we
get
\begin{equation*}
  \bigl| U + \normCDF^{-1}(1-p) \bigr| 
  \;=\;
  \left| U - \normCDF^{-1}(p) \right|
  \;\stackrel{\mbox{\scriptsize d}}{=}\;
  \left| U + \normCDF^{-1}(p) \right|,    
\end{equation*}
where $\stackrel{\mbox{\scriptsize d}}{=}$ denotes an equality in
distribution. Therefore $G_{\kappa,\delta}(1-p) =
G_{\kappa,\delta}(p)$.
  
\medbreak\noindent 
b)\ Let $S_p = \max\bigl( 0, \kappa^\delta - \bigl| \normCDF^{-1}(p) +
U \bigr| \bigr)$. Straightforward computations show that $t \mapsto
\P\left( \left| t + U \right| \le v \right)$ is strictly decreasing
to~$0$ on~$\left[ 0, +\infty \right[$, for all~$v > 0$. As a
consequence, $p \mapsto \P\bigl( S_p < s \bigr)$ is strictly
increasing to~$1$ on $\left[ 1/2, 1 \right[$, for all $s \in \bigl] 0,
\kappa^\delta \bigr[$. Therefore, $G_{\kappa, \delta}$ is strictly
decreasing on $\left[ 1/2, 1 \right[$ and tends to zeros when $p \to
1$. The other assertions then follow from~a).

\medbreak\noindent
c) Recall that $\xi(x) \sim \Ncal\bigl( \xihat_n(x), \sigma_n^2(x)
\bigr)$ under~$\P_n$. Therefore $U := \bigl( \xi(x) - \xihat_n(x)
\bigr) / \sigma_n(x) \sim \Ncal(0,1)$ under~$\P_n$, and the result
follows by substitution in~\eqref{RanjBichLikeCriteria}.

\medbreak\noindent The closed-form expressions of Ranjan et al.'s and
Bichon and al.'s criteria (assertions~\ref{prop:cf-bichon})
and~\ref{prop:cf-ranjan})) is established in the following sections.

\subsection{A preliminary decomposition common to both criteria}

\newcommand \Gad {G_{\kappa,\delta}}

Recall that $t = \normCDF^{-1}(1-p)$, $t^+ = t + \kappa$ and $t^- = t - \kappa$. Then,
\begin{align}
  \Gad(p) 
  \;=\;&\ 
  \Gad(1-p) \;=\; \EE\left( \max\Bigl( 
    0, \kappa^\delta - \bigl| t - U \bigr|^\delta 
    \Bigr) \right)
  \nonumber \\
  \;=\;&\ 
  \int_{\kappa^\delta - \left| t - u \right|^\delta \ge 0} 
  \left( \kappa^\delta - \left| t - u \right|^\delta \right)\,
  \normPDF(u)\, \du
  \nonumber \\
  \;=\;&\ 
  \int_{t^-}^{t^+}
  \left( \kappa^\delta - \left| t - u \right|^\delta \right)\,
  \normPDF(u)\, \du
  \nonumber \\ 
  \;=\;&\
  \kappa^\delta \left( \normCDF(t^+) - \normCDF(t^-) \right)
  \,-\,
  \underbrace{
    \int_{t^-}^{t^+} \left| t - u \right|^\delta \normPDF(u)\, \du
  }_{\text{Term $A$}}.
  \label{eq:G-two-terms-decomp}
\end{align}

\noindent
The computation of the integral~$A$ will be carried separately in the
next two sections for~$\delta = 1$ and~$\delta = 2$. For this purpose,
we shall need the following elementary results:
\begin{align}
  \int_a^b u \normPDF(u) \du \;=\;&  \normPDF(a) - \normPDF(b)
  \label{cdv2} \,,\\
  \int_a^b u^2 \normPDF(u) \du \;=\;&  a \normPDF(a) - b\normPDF(b) + \normCDF(b) - \normCDF(a) \,.
  \label{cdv3}
\end{align}

\subsection{Case $\delta = 1$}

Let us compute the value~$A_1$ of the integral~$A$ for~$\delta = 1$:
\begin{align}
  A_1 \;=\;&
  \int_{t^-}^{t^+} \left| t-u \right| \normPDF(u) \du 
  \;=\;
  \int_{t^-}^{t} (t-u) \normPDF(u) \du 
  + \int_{t}^{t^+} (u-t) \normPDF(u) \du
  \nonumber \\
  \;=\;&
  t \left( \int_{t^-}^{t} \normPDF(u)\, \du - \int_{t}^{t^+} \normPDF(u)\, \du\right)
  - \int_{t^-}^{t} u \normPDF(u)\, \du + \int_{t}^{t^+} u
  \normPDF(u)\, \du 
  \nonumber \\
  \;=\;&
  t \left( 2\normCDF(t) - \normCDF(t^-) - \normCDF(t^+) \right)
  \,+\,
  2\normPDF(t) - \normPDF(t^-) - \normPDF(t^+) \,,
\label{eq:comp-bichon-termA}
\end{align}
where \eqref{cdv2} has been used to get the final result. Plugging
\eqref{eq:comp-bichon-termA} into~\eqref{eq:G-two-terms-decomp}
yields~\eqref{eq:bichon-explicit}.

\subsection{Case $\delta = 2$}

Let us compute the value~$A_2$ of the integral~$A$ for~$\delta = 2$:
\begin{align}
  A_2 \;=\;& 
  \int_{t^-}^{t^+} (t-u)^{2} \normPDF(u)\, \du 
  \nonumber \\
  \;=\;&
  t^2 \int_{t^-}^{t^+} \normPDF(u)\, \du 
  - 2t \int_{t^-}^{t^+} u \normPDF(u)\, \du  
  + \int_{t^-}^{t^+} u^2 \normPDF(u)\, \du
  \nonumber \\
  \;=\;&
  t^2 \left( \normCDF(t^+) - \normCDF(t^-) \right)
  \,-\, 2t \left( \normPDF(t^-) - \normPDF(t^+) \right)
  \nonumber \\
  & +\, t^- \normPDF(t^-) - t^+ \normPDF(t^+) 
  \,+\, \normCDF(t^+) - \normCDF(t^-) \,,
  \label{eq:comp-ranjan-termA}
\end{align}
where \eqref{cdv2} and (\ref{cdv3}) have been used to get the final
result. Plugging \eqref{eq:comp-bichon-termA} into
\eqref{eq:G-two-terms-decomp} yields \eqref{eq:ranjan:explicit}.

% BST file found at http://links.tedpavlic.com/bst/spmpscinat.bst
\bibliographystyle{spmpscinat}
\bibliography{StatAndComp}
\end{document}